\documentclass[prd,twocolumn,showpacs,floatfix,superscriptaddress]{revtex4}
\usepackage{graphicx}
\usepackage{subfigure}
\usepackage{epsf}
\usepackage{bm}
\usepackage{amsmath}
\usepackage{amsfonts}
\usepackage{amssymb}
\usepackage{graphicx}
\usepackage{tabularx}
\usepackage{color}%
\providecommand{\U}[1]{\protect\rule{.1in}{.1in}}
\newcommand{\be}{\begin{equation}}
\newcommand{\ee}{\end{equation}}

\newcommand{\mincir}{\raise
-3.truept\hbox{\rlap{\hbox{$\sim$}}\raise4.truept\hbox{$<$}\ }}
\newcommand{\magcir}{\raise
-3.truept\hbox{\rlap{\hbox{$\sim$}}\raise4.truept\hbox{$>$}\ }}

\providecommand{\U}[1]{\protect\rule{.1in}{.1in}}

\usepackage{tikz,xcolor,hyperref}

\definecolor{lime}{HTML}{A6CE39}
\DeclareRobustCommand{\orcidicon}{%
	\begin{tikzpicture}
	\draw[lime, fill=lime] (0,0) 
	circle [radius=0.16] 
	node[white] {{\fontfamily{qag}\selectfont \tiny ID}};
	\draw[white, fill=white] (-0.0625,0.095) 
	circle [radius=0.007];
	\end{tikzpicture}
	\hspace{-2mm}
}

\foreach \x in {A, ..., Z}{%
	\expandafter\xdef\csname orcid\x\endcsname{\noexpand\href{https://orcid.org/\csname orcidauthor\x\endcsname}{\noexpand\orcidicon}}
}

\begin{document}

\title{Scalar Field cosmology from a Modified Poisson Algebra}
\author{Genly Leon\orcidA{}}
\email{genly.leon@ucn.cl}
\affiliation{Departamento de Matem\'{a}ticas, Universidad Cat\'{o}lica del Norte, Avda.
Angamos 0610, Casilla 1280 Antofagasta, Chile}
\affiliation{Institute of Systems Science, Durban University of Technology, PO Box 1334,
Durban 4000, South Africa}

\author{Alfredo D. Millano\orcidC{}}
\email{alfredo.millano@alumnos.ucn.cl}
\affiliation{Departamento de Matem\'{a}ticas, Universidad Cat\'{o}lica del Norte, Avda.
Angamos 0610, Casilla 1280 Antofagasta, Chile}

\author{Andronikos Paliathanasis\orcidB{}}
\email{anpaliat@phys.uoa.gr}
\affiliation{Institute of Systems Science, Durban University of Technology, PO Box 1334,
Durban 4000, South Africa}
\affiliation{Departamento de Matem\'{a}ticas, Universidad Cat\'{o}lica del Norte, Avda.
Angamos 0610, Casilla 1280 Antofagasta, Chile}

\begin{abstract}
We investigate the phase space of a scalar field theory obtained by minisuperspace deformation. We consider quintessence or phantom scalar fields in the action which arise from minisuperspace deformation on the Einstein-Hilbert action. We use a modified Poisson Algebra where Poisson brackets are the $\alpha$-deformed ones and are related to the Moyal-Weyl star product. We discuss early and late-time attractors, and we reconstruct the cosmological evolution. We show that the model can have $\Lambda$CDM model as a future attractor if we initially consider a massless scalar field without a cosmological constant term. 

\end{abstract}
\date{\today}
\maketitle

\section{Introduction}

Cosmological observations indicate that the universe has gone through two
acceleration phases \cite{Teg, Kowal, Komatsu,planck}, an early acceleration
phase known as inflation and the present acceleration phase. The source of
the cosmic acceleration is unknown. In the context of General Relativity, cosmic
acceleration occurs when the cosmic fluid is dominated by a matter source
known as dark energy (DE) with the property to have a negative value of the
equation of state (EoS) parameter.

The cosmological constant $\Lambda,$ leads to the $\Lambda$-cosmology is
indeed the simplest candidate for DE; however, it suffers from two
problems, the fine-tuning and the coincidence problems
\cite{Padmanabhan1,Weinberg1}. Furthermore, the detailed analysis of the
recent cosmological observations shows that $\Lambda$-cosmology cannot solve
tensions arise from the statistical analysis of the data, such as the
$H_{0}$-tension \cite{h0ten}. There are various DE alternatives to
the cosmological constant, which has been proposed to
overpass the mentioned above problems; see, for instance
\cite{Ratra88,Lambdat,Bas09c,Wetterich:1994bg,Caldwell98,Brax:1999gp,Caldwell,LSS08,Brookfield:2005td,Ame10},
and references therein.

Scalar fields play a significant role in the description of cosmic
acceleration. Indeed, introducing a scalar field in the field
equations provides new degrees of freedom in the gravitational dynamics 
that to provide acceleration effects. The most straightforward mechanism for describing
the early acceleration phase of the universe, that is, of the
inflationary epoch, is that of the inflaton field
\cite{ref1,ref2,ref3,ref4,ref5,ref6,ref7,ref8}. During inflation,
\cite{Aref1,guth} the scalar field dominates the cosmological dynamics and
provides the antigravity effects. Similarly, for the description of the
late-time acceleration  \cite{Dolgov82}, a tracker
scalar field can be introduced \cite{Caldwell98}, which roles down
the potential energy $V(\phi)$ such that to have DE effects
\cite{Jassal,SR,SVJ}. Another novelty of the scalar fields is that they can
reproduce various DE alternatives such as the Chaplygin gas and
others \cite{alt1,alt2}.

In quintessence scalar field cosmology \cite{Ratra88}, the EoS
parameter of the scalar field is constrained to the range
$\left\vert w_{\phi}\right\vert \leq1$, where $w_{\phi}=1$ correspond to a
stiff fluid where only the kinetic part of the scalar field dominates, while
the limit $w_{\phi}=-1$ correspond to the case where only the scalar field
potential dominates, leading to $\Lambda$-cosmology. Recall that acceleration is
occurred when $-1\leq w_{eff}<-1/3.~$There is a family of scalar field models,
known as phantom scalar fields where $w_{\phi}\,$\ can cross the limit $-1$
and take smaller values, which is possible, for example, when there exists a negative kinetic
energy \cite{ph1,ph2,ph3,ph4}.

During the very early stages of the universe, we expect that quantum effects
play an important role in cosmic evolution. Until now, there is not a
unique theory of quantum gravity; that is why various approaches have been
considered in the literature by various groups
\cite{MB01,Niem01,Kempf01,KN01,Amjad1,Amjad2,an1,an2,mant}. String theory,
double special relativity and generalised uncertainty principle require the
existence of a minimum length scale of the order of the Planck length
$l_{\mathrm{pl}}$
\cite{Mukhi:2011zz,KowalskiGlikman:2004qa,AmelinoCamelia:2010pd,Bekenstein1,
Bekenstein2,Maggiore,Giacomini:2020zmv,Paliathanasis:2021egx}. As a result of
the modification of the Heisenberg uncertainty in the latter approaches a
deformation parameter is introduced, which leads to the deformation of the
coordinate representation of the operators of the momentum
position, that is, to a deformation of the Poisson algebra \cite{ss1}.

In \cite{Perez-Payan:2011cvf}, the phase space for the cosmological dynamics in
quintessence cosmology was modified by a deformed Poisson algebra among the
coordinates and the canonical momenta. The main result was that the
deformation parameter is related to the accelerating scale factor provided
by the deformed Poisson algebra in the absence of a cosmological constant. A
similar result was determined recently in \cite{bat1} and in the case of a
phantom scalar field.

The Moyal-Weyl star product provides a simple prescription for
constructing Non-commutative field theories on the noncommutative manifold \cite{Perez-Payan:2011cvf}
with $[\hat{x}^{\mu},\hat{x}^{\nu}]=i\theta^{\mu\nu}$. One simply replaces all the point-wise products in ordinary field theory with one of the star products.
For example, the Non-commutative action for a real massless scalar field
$\Phi$ in four dimensions is
\begin{equation}
S_{\phi}=\frac{1}{2}\int d{}^{4}xD^{\mu}\Phi\star_{\alpha}D_{\mu}\Phi,
\end{equation}
where the ordinary derivative $\partial_{\mu}$ appearing in the commutative scalar field action is replaced by the non-commutative covariant derivative
$D_{\mu}$, so the action is invariant under the non-commutative gauge transformation.

In this paper, we are interested in studying the effects of the deformed
Poisson algebra in the cosmological evolution. Specifically, we
perform a detailed analysis of the phase-space to investigate the
existence of equilibrium points and reconstruct the cosmological parameters' evolution. Such analysis provides important information about
the theory's viability and can give us important results for the nature
of the deformation parameter. For this analysis, one can introduce auxiliary
variables which transform the cosmological equations into an autonomous
dynamical system
\cite{din1,din2,din3,din4,din5,din6,din7,din8,din9,din10,din11,din12,din13,din14,din15}. Hence, we obtain a system of the form  $\textbf{X}'=\textbf{f(X)}$, where $\textbf{X}$ is the
column vector of the auxiliary variables, $\textbf{f(X)}$ is an autonomous vector field, and the derivative is with respect to a logarithmic time-scale. The stability analysis comprises several steps. First, the critical points $\bf{X_c}$ are extracted under the requirement  of $\bf{X}'=0$. Then, one consider linear perturbations around $\bf{X_c}$ as $\bf{X}=\bf{X_c}+\bf{U}$, with $\textbf{U}$ the column vector of the  auxiliary variable's perturbations. Therefore, up
to first order we obtain $\textbf{U}'={\bf{\Xi}}\cdot \textbf{U}$, where  the matrix ${\bf {\Xi}}$ contains coefficients of the perturbed equations. Finally, the type stability of each hyperbolic critical point is determined by the eigenvalues of ${\bf {\Xi}}$. That is, the point is stable (unstable) if the reals parts of the eigenvalues are negative (positive) or a saddle point
if the eigenvalues have real parts with different signs. 

The structure of the paper is as follows.  In Section II, we introduce the modified Poisson algebra. In Section III, we derive the modified field equations in the case of scalar field cosmology in an isotropic and homogeneous spatially flat universe. Sections IV and V include the main results of this study, where we present the detailed analysis of the phase space for the modified field equations. Finally, in Section VI, we summarise our results and draw conclusions.

\section{Modified Poisson Algebra}

We consider the modified Poisson Algebra \cite{Perez-Payan:2011cvf}
\begin{align}
    & \{x_1, x_j\}_\alpha = \theta _{i j}, \\ 
    & \{p_i, p_j\}_\alpha = \beta _{i j}, \\ 
    & \{x_i, p_j\}_\alpha = \delta _{i j} +\sigma_{i j},
\end{align}
where the Moyal-Weyl brackets are defined through the relation
\begin{equation}
    \{f, g\}_\alpha = f \star_{\alpha} g - g \star_{\alpha} f
\end{equation}
in which the product between $f$ and $g$ is substituted by the Moyal-Weyl star product
\begin{equation}
    (f \star_{\alpha} g)= \exp\left[\frac{1}{2} \alpha^{a b} \partial_a^{(1)} \partial_b^{(2)}\right] f(x_1) g(x_2)\Big|_{x_1=x_2=x}, 
\end{equation}
such that 
\begin{equation}
    \alpha= \begin{pmatrix}
\theta_{i j}&  \delta_{i j} + \sigma_{i j}\\
- \delta_{i j}- \sigma_{i j} & \beta_{i j}
\end{pmatrix},
\end{equation}
where $\theta_{i j}$ and $\beta_{i j}$ are $2\times 2$ antisymmetric matrices indicating the non-commutativity in the coordinates and momenta, respectively.  Particular deformations 
\begin{equation}
    \theta_{i j}= - \theta \epsilon_{i j}, \; \beta_{i j}= \beta \epsilon_{i j},
\end{equation}
where $\epsilon_{i j}$ is the two-indices Levi-Civita symbol, are considered.

 By removing  the sub-index in  $\star_{\alpha}$, the $\star$-Friedman equations can be derived for $\star$-FLRW metric as follows \cite{deVegvar:2020now}.

\begin{equation}
R_{\mu \nu }-{\frac {1}{2}}R \star g_{\mu \nu }+\Lambda g_{\mu \nu }=\kappa T_{\mu \nu }
\end{equation}
with energy-momentum tensor
\begin{equation}
 T_{\mu \nu } = p \star g_{\mu \nu} +(\rho+p)\star U_{\mu} \star U_{\nu}
\end{equation}
where $U_\mu= \delta_\mu^{0}$ is the co-moving observer, $p$ and $\rho$ are the total pressure and fluid energy 3-density respectively. 

To avoid complexities of $\star$-algebras, one may consider the field equations arising from the point-like action for a scalar field with action \cite{Basilakos:2011rx}
\begin{equation}
    S= \int d t N \left(-3\frac{a {\dot{a}}^2}{N^2}\right) + \frac{1}{2} \int d t N a^3 \left(\epsilon \frac{{\dot{\phi}}^2}{N^2}- 2 V(\phi)\right).
\end{equation}
We define the point-like Lagrangian \cite{Basilakos:2011rx}
\begin{equation}
 \mathcal{L}(N, a, \phi, \dot{a}, \dot{\phi}):=   \frac{1}{N} \left(-3 a \dot{a}^2 +\frac{1}{2}a^3 \epsilon {\dot{\phi}}^2\right) - a^3 N V(\phi),\label{ns.11} 
\end{equation}
while for simplicity we consider a constant potential $ V(\phi)=\tilde{\Lambda}$. Sign $\epsilon=1$ corresponds to quintessence, and sign $\epsilon=-1$ corresponds to the phantom field. 

Variation with respect to $\{N, a, \phi\}$ and the replacement $N=1$ after variation, we obtain the Euler-Lagrange equations 
\begin{align}
& \frac{1}{2} \left(6 a  {\dot{a}}^2-a ^3 \left(2 \tilde{\Lambda}  +\epsilon  {\dot{\phi}}^2\right)\right)=0,\\
& \frac{3}{2} \left(4 a 
   \ddot{a}+2 {\dot{a}}^2+a ^2 \left(\epsilon  {\dot{\phi}}^2-2 \tilde{\Lambda} \right)\right)=0,\\
& -\epsilon  a ^2 \left(3 {\dot{a}} {\dot{\phi}}+a    {\ddot{\phi}}\right)=0. 
\end{align}
Introducing the Hubble parameter $H=\dot{a}/a$ the previous equations can be written as \cite{Basilakos:2011rx}
\begin{align}
   & 3 H ^2= \tilde{\Lambda} +\frac{1}{2} \epsilon  {\dot{\phi}}^2,\\
   & 2 \dot{H} =-3 H ^2+\tilde{\Lambda} -\frac{1}{2} \epsilon  {\dot{\phi}}^2,\\
   &{\ddot{\phi}}+ 3  H  {\dot{\phi}}=0.
\end{align}

 For the Lagrangian function \eqref{ns.11} we define the generalised momenta by
$p_{i}=\frac{\partial {\mathcal {L}}}{\partial {\dot {q}}^{i}}$, where $q^i\in\{a, \phi\}$, $p_i\in\{p_a, p_\phi\}$,  namely
\begin{align}
p_{a}  &  \equiv -\frac{6 a \dot{a}}{N}, \; 
p_{\phi}  \equiv \frac{\epsilon  a^3 \dot{\phi}}{N}.
\label{ns.18}%
\end{align}
Hence, we can introduce the Hamiltonian function  
$\mathcal{H}=p_{a}\dot{a}+p_{\phi}\dot{\phi}-\mathcal{L}$, which is written as
\begin{equation}
\mathcal{H}=    N \left(-\frac{{p_a}^2}{12 a}+\frac{ \epsilon {p_{\phi}}^2
  }{2 a^3}+\tilde{\Lambda}  a^3\right)
\end{equation}

We define the canonical coordinates  \cite{Basilakos:2011rx}
\begin{equation}
  x= \lambda^{-1} (\epsilon a)^{3/2} \sinh(\sqrt{\epsilon}\lambda \phi), \; 
y= \lambda^{-1} a^{3/2} \cosh(\sqrt{\epsilon}\lambda \phi),
\end{equation}
with inverse 
\begin{equation}
 a=   \left(\lambda  y  \sqrt{1-\frac{\epsilon  x^2}{y^2}}\right)^{2/3}, \; 
\phi=    \frac{\tanh ^{-1}\left( \sqrt{\epsilon }
   x/y\right)}{\lambda  \sqrt{\epsilon }}, 
\end{equation}
where $\lambda^{-1}= \sqrt{8/3}$, and we consider the simpler case where the matter content is an ordinary ($\epsilon=+1$) or a phantom ($\epsilon=-1$) scalar field in action.
Then, \eqref{ns.11} becomes 
\begin{equation}
 \mathcal{L}(N, a, \phi, \dot{a}, \dot{\phi}):=   \frac{1}{N} \left(-3 a \dot{a}^2 +\frac{1}{2}a^3 \epsilon {\dot{\phi}}^2\right) - a^3 N V(\phi). \label{ns.11b} 
\end{equation}
Generalised momenta are given by
\begin{align}
    P_x= \frac{\epsilon \dot{x}}{N}, \; 
    P_y=-\frac{\dot{y}}{N}. \label{ns.18b}
\end{align}
Hence, the problem can be formulated from the canonical Hamiltonian 
\begin{equation}
    \mathcal{H}_c = \epsilon N \left(\frac{1}{2} P_x^2 + \frac{\omega^2}{2} x^2\right) -  N \left(\frac{1}{2} P_y^2 + \frac{\omega^2}{2} y^2\right),
\end{equation}
where $\omega^2= -3 \tilde{\Lambda}/4$, and we use the comoving frame $N=1$. For the choice $\epsilon=+1$, see related work  \cite{Perez-Payan:2011cvf}.

We have the evolution Eqs. for $(\dot{x}, \dot{y})$ as given by \eqref{ns.18b}: 
\begin{equation}
    \dot{x}= \epsilon P_x, \; \dot{y}= -P_y
\end{equation}
Hamilton's equations ${\dot {p}}_{i} 
= - {\frac {\partial {\mathcal {H}}}{\partial q^{i}}}$, where  $\mathcal{H}=\mathcal {H}_c$ and $q^i\in\{x, 
y\}$, $p_i\in\{P_x, P_y\}$, lead to  
\begin{equation}
    \dot{P_x}= -\epsilon \omega^2 x, \;    \dot{P_y}=  \omega^2 y.
\end{equation}
which lead to the following equations for $\epsilon=\pm 1$,
\begin{equation}
    \ddot{x}+ \omega^2 x=0, \;      \ddot{y}+ \omega^2 y=0,
\end{equation}
with conserved quantity 
\begin{equation}
    \tilde{y}^2-\epsilon \tilde{x}^2=1, \; (\tilde{x},\tilde{y})= \lambda a^{-3/2} \cdot (x,y).
\end{equation}
By definition $\omega^2=-3/4\tilde{\Lambda}$, so the solutions are 
\begin{align}
& x(t)= c_1 e^{\frac{1}{2} \sqrt{3} \sqrt{\tilde{\Lambda} }    t}+c_2 e^{-\frac{1}{2} \sqrt{3} \sqrt{\tilde{\Lambda} } t},\\
& y(t)= c_3   e^{\frac{1}{2} \sqrt{3} \sqrt{\tilde{\Lambda} } t}+c_4 e^{-\frac{1}{2}   \sqrt{3} \sqrt{\tilde{\Lambda} } t}.
\end{align}
Then,
\begin{align}
\phi(t)& =   \frac{\coth ^{-1}\left(\frac{\epsilon ^{3/2} \left(c_3
   e^{\sqrt{3} \sqrt{\tilde{\Lambda} } t}+c_4\right)}{c_1 e^{\sqrt{3}
   \sqrt{\tilde{\Lambda} } t}+c_2}\right)}{\lambda  \sqrt{\epsilon
   }},\\
a(t)& =  e^{- \sqrt{\frac{\tilde{\Lambda}}{3} } t}\left(\lambda 
   \left(c_3 e^{\sqrt{3} \sqrt{\tilde{\Lambda} } t}+c_4\right) \right)^{2/3} \nonumber \\
   & \times \left(
 {1-\frac{\epsilon  \left(c_1 e^{\sqrt{3} \sqrt{\tilde{\Lambda} }
   t}+c_2\right){}^2}{\left(c_3 e^{\sqrt{3} \sqrt{\tilde{\Lambda} }
   t}+c_4\right){}^2}}\right)^{1/3}, 
\end{align}
such that 
\begin{align}
& \phi \sim \frac{\ln \left(1-\frac{2 c_1}{c_1-c_3 \epsilon
   ^{3/2}}\right)}{2 \lambda  \sqrt{\epsilon }}, \;  a \sim c_3{}^{2/3}
   \lambda ^{2/3} \sqrt[3]{1-\frac{c_1{}^2 \epsilon }{c_3{}^2}}
   e^{\frac{\sqrt{\tilde{\Lambda} } t}{\sqrt{3}}}
\end{align}
as $t\rightarrow \infty$. That is, a de Sitter solution is obtained.

The elements of the new configuration space, $(x , y)$, and
their conjugate momenta fulfil the following commutation relations based on the Poisson bracket:
\begin{align}
    \{x_k, x_j\} = 0, \{P_{x_k}, P_{x_j}\} = 0, \{x_k, P_{x_j}\} = \delta_{k j}
\end{align}
where $k$ and $j$ can take $1$ and $2$, that is $(x_1, x_2) = (x, y)$ and
$\delta_{k j}$ is the usual Kronecker delta.

To obtain a modified scenario, we take classical phase space variables $(x, y, P_x, P_y)$ and perform the transformation (see related work \cite{Perez-Payan:2011cvf}): 
\begin{equation}
    \begin{pmatrix}
    \hat{x}\\
    \hat{y}
    \end{pmatrix}=     \begin{pmatrix}
    x\\
   y
    \end{pmatrix} +\frac{\theta}{2}   \begin{pmatrix}
   P_y\\
   -P_x
    \end{pmatrix} 
\end{equation}
and 
\begin{equation}
    \begin{pmatrix}
    \hat{P}_x\\
    \hat{P}_y
    \end{pmatrix}=   \frac{\beta}{2}   \begin{pmatrix}
  -y\\
   x
    \end{pmatrix} + \begin{pmatrix}
  P_x\\
  P_y
    \end{pmatrix} 
\end{equation}
the modified Poisson Algebra is given by 
\begin{equation}
    \{\hat{y}, \hat{x}\}= \theta, \;   \{\hat{P}_y, \hat{P}_x\}= \beta
\end{equation}
and
\begin{equation}
    \{\hat{x}, \hat{P}_x\}= 1+ \sigma, \;    \{\hat{y}, \hat{P}_y\}= 1+ \sigma,
\end{equation}
where $\sigma= \theta \beta/4$. Now, we change notation $ \{\hat{x}, \hat{y}, \hat{P}_x, \hat{P}_y\}$ to  $ \{x, y, p_x, p_y\}$.

The modified Hamiltonian will be 
\begin{equation}
    \mathcal{H}_{mod.}=  \frac{1}{2}\epsilon p_x^2 -    \frac{1}{2} p_y^2 -\frac{\omega_1^2}{2} \left( x p_y + \epsilon y p_x\right) + \frac{\omega_2^2}{2} ( \epsilon x^2 - y^2),
\end{equation}
where 
\begin{align}
    & p_x=\dot{x}+\frac{1}{2}\omega_1y, \, p_y=\frac{-x \omega_1^2 \epsilon -2 \dot{y}}{2 \epsilon },
\end{align} and 
we define the parameters 
\begin{align}
    & \omega_1^2= \frac{4 \beta -4 \epsilon \omega^2 \theta}{4 - \epsilon \omega^2 \theta^2}, \; \omega_2^2=\frac{4 \omega^2 - \epsilon  \beta^2}{4 - \epsilon \omega^2 \theta^2}, \\
    & \Lambda=-\frac{4 \left((\beta -1) \beta  \epsilon +(\theta -4)
   \omega ^2\right)}{3 \left(\theta ^2 \omega ^2 \epsilon
   -4\right)}. \label{param}
\end{align}

If $\omega=0$, the latter definitions are
\begin{align}
    & \omega_1^2= \beta, \; \omega_2^2=-\frac{\epsilon \beta^2}{4}, \;
     \Lambda=\frac{(\beta-1)\beta \epsilon}{3}. \label{param2}
\end{align}
We can infer from these that the cosmological constant term is introduced from the modification of the Poisson algebra if our initial model does not include a cosmological constant term. 
The equations of motion derived from $ \mathcal{H}_{mod.}$ are
\begin{equation}
    \ddot{x} + \omega_1^2 \dot{y} - \frac{3}{4}  \Lambda x=0, \label{eq.1}
\end{equation}
and 
\begin{equation}
    \ddot{y} + \epsilon \omega_1^2 \dot{x} -\frac{3}{4}  \Lambda y=0, \label{eq.2}
\end{equation}
These equations have solutions 
\begin{widetext}
\begin{align}
 x(t) & = c_1 \cosh \left(\frac{t {\omega_1}^2}{2}\right) \cosh \left(\frac{1}{2} t \sqrt{3 \Lambda
   +{\omega_1}^4}\right)-c_3 \sinh \left(\frac{t
   {\omega_1}^2}{2}\right) \cosh \left(\frac{1}{2} t
   \sqrt{3 \Lambda +{\omega_1}^4}\right) \nonumber\\
   & +\frac{\sinh
   \left(\frac{1}{2} t \sqrt{3 \Lambda +{\omega_1}^4}\right) \left(\left(c_3 {\omega_1}^2+2 c_2\right)
   \cosh \left(\frac{t {\omega_1}^2}{2}\right)-\left(c_1
   {\omega_1}^2+2 c_4\right) \sinh \left(\frac{t
   {\omega_1}^2}{2}\right)\right)}{\sqrt{3 \Lambda
   +{\omega_1}^4}},\\
   y(t) & = c_3 \cosh \left(\frac{t
   {\omega_1}^2}{2}\right) \cosh \left(\frac{1}{2} t
   \sqrt{3 \Lambda +{\omega_1}^4}\right)-c_1 \sinh
   \left(\frac{t {\omega_1}^2}{2}\right) \cosh
   \left(\frac{1}{2} t \sqrt{3 \Lambda +{\omega_1}^4}\right) \nonumber \\
   & +\frac{\sinh \left(\frac{1}{2} t \sqrt{3 \Lambda
   +{\omega_1}^4}\right) \left(\left(c_1 {\omega_1}^2+2 c_4\right) \cosh \left(\frac{t {\omega_1}^2}{2}\right)-\left(c_3 {\omega_1}^2+2 c_2\right)
   \sinh \left(\frac{t {\omega_1}^2}{2}\right)\right)}{\sqrt{3 \Lambda +{\omega_1}^4}}.
\end{align}
\end{widetext}
Some solutions of this form have been found before in the literature, e.g., \cite{Paliathanasis:2014yfa, Paliathanasis:2018vru}. 

\section{Modified Friedmann equations}

Equations \eqref{eq.1} and \eqref{eq.2}
are equivalent to 
\begin{align}
   &   {\ddot{a}} = -\frac{{\dot{a}} ^2}{2 a }-\frac{1}{6} a  \left(4 \lambda  \epsilon  {\dot{\phi}}  \left(\lambda  {\dot{\phi}} +{\omega_1}^2\right)+4 {\omega_2}^2+ \omega_{1}^2 \epsilon \right),\\
   &  {\ddot{\phi}}= -\frac{3 {\dot{a}}  \left(2 \lambda  {\dot{\phi}} +{\omega_1}^2\right)}{2 \lambda  a }
\end{align}
with first integral 
\begin{equation}
-3 H ^2+\frac{1}{2} \epsilon  {\dot{\phi}} ^2-\frac{4 {\omega_2}^2}{3}-\frac{{\omega_1}^2 \epsilon }{3}=0.
\end{equation}
where $H=\dot{a}/a$ is the Hubble parameter.  

\subsection{Vacuum case}

Using the reparameterization \eqref{param}, the modified Friedman equation reads 
\begin{equation}
    3 H^2 = \frac{1}{2} \epsilon {\dot{\phi}}^2   +\Lambda. \label{FRW1}
\end{equation}

The modified Klein-Gordon equation is 
\begin{equation}
    \ddot{\phi} = - H  \left(\sqrt{6}   \omega _{1}^2 + 3 \dot{\phi}\right)  \label{FRW2}
\end{equation}
Raychaudhuri equation is 
\begin{equation}
   \dot{H}=  -\frac{3 H ^2}{2}-\frac{  \epsilon  \omega_{1}^2
 \dot{\phi}}{\sqrt{6}}-\frac{1}{4} \epsilon  {\dot{\phi}}^2 + \frac{\Lambda}{2},
\end{equation}
Alternatively, by removing $H^2$ using \eqref{FRW1}, we obtain, 
\begin{equation}
   \dot{H}=  -\frac{ \epsilon\omega_{1}^2 \dot{\phi}}{\sqrt{6}}-\frac{1}{2} \epsilon  {\dot{\phi}}^2.  \label{FRW3}
\end{equation}
With the definitions 
\begin{equation}
\rho_\phi=    \frac{1}{2} \epsilon {\dot{\phi}}^2 
+\Lambda, \; 
P_\phi =   \frac{1}{2} \epsilon {\dot{\phi}}^2 +\sqrt{\frac{2}{3}} \omega_{1}^2 \epsilon  \dot{\phi}-\Lambda, \label{sfdensities}
\end{equation}
the Klein-Gordon equation can be written as the  conservation equation
\begin{equation}
    \dot{\rho}_\phi + 3 H(\rho_\phi + P_\phi)= 0.
\end{equation}
Moreover, we define the effective EoS parameter of $\phi$ as 
\begin{equation}
w_\phi: = \frac{P_\phi}{\rho_\phi}=  \frac{   \epsilon {\dot{\phi}}^2 +2\sqrt{\frac{2}{3}} \omega_{1}^2 \epsilon  \dot{\phi}-2 \Lambda}{\epsilon {\dot{\phi}}^2 +2\Lambda}. \label{EoS_phi}
\end{equation}

\subsection{Including matter}

The Friedman equation reads 
\begin{equation}
    3 H^2 = \frac{1}{2} \epsilon {\dot{\phi}}^2  + \rho_m  +\Lambda.
\end{equation}
The modified Klein-Gordon equation is 
\begin{equation}
    \ddot{\phi} = -  H  \left(\sqrt{6}   \omega _{1}^2 + 3 \dot{\phi}\right)
\end{equation}
that can be written using \eqref{sfdensities} as 
\begin{equation}
    \dot{\rho}_\phi + 3 H(\rho_\phi + P_\phi)= 0.
\end{equation}
We have the matter conservation equation
\begin{equation}
    \dot{\rho}_m +3 H (1+ w_m) \rho_m=0.
\end{equation}
Raychaudhuri equation is 
\begin{equation}
  \dot{H}= -\frac{ \omega_{1}^2 \epsilon\dot{\phi}}{\sqrt{6}}-\frac{1}{2} \epsilon  {\dot{\phi}}^2 -\frac{1}{2}\left( w_m+1\right)  \rho_{m}.
\end{equation}

\section{Dynamical systems analysis in vacuum case}
In this Section, we proceed with the analysis of the phase space for the modified cosmological field equations. In order to perform such an analysis we define dimensionless variables in the Hubble-normalization approach, that is, 
\begin{equation}
    \Sigma_\phi=\frac{\dot{\phi}}{\sqrt{6}H}, \; {\Sigma}=  \frac{\omega_1^2}{H}, 
\end{equation}
which satisfies the constraint equation
\begin{equation}
     \Sigma_\phi^2 \epsilon + \mu  {\Sigma}^2 =1, \label{const1}
\end{equation}
where we have introduced the constant $\mu= \Lambda/(3\omega_1^4)$, 
or, alternatively
\begin{equation}
     \Sigma_\phi^2 \epsilon + \Omega_\Lambda =1.
\end{equation}
whereby convenience, we define the fractional energy density of $\Lambda$ as 
\begin{equation}
    \Omega_\Lambda:=\frac{\Lambda}{3 H^2}= \mu {\Sigma}^2,\; \mu>0. \label{Lambda_eq}
\end{equation}
Thus, the dynamical system \eqref{FRW1}, \eqref{FRW2} and\eqref{FRW3} can be written as a dynamical system as
\begin{align}
 & \Sigma_\phi^{\prime}=   (3 \Sigma_\phi+{\Sigma}) \left(\Sigma_\phi^2 \epsilon -1\right), \label{syst1a}\\
 & {\Sigma}^{\prime}= \epsilon \Sigma_\phi {\Sigma}   (3
   \Sigma_\phi+{\Sigma}), \label{syst1b}
\end{align}
where we have introduced the new time derivative $f^{\prime}=H^{-1} \dot{f}$.

\begin{table}[!t]
    \centering
    \begin{tabular}{|c|c|c|c|c|c|}
    \hline
  \text{Label} & \text{Existence} & \text{Coordinates}\; $(\Sigma_{\phi},  \Sigma)$ & \text{Eigenvalues }& \text{Stability}\\\hline   
  $P_{1,2}$ & $\epsilon=+1$ &$(\pm 1,0)$& $\{3, 6\}$ & \text{Unstable}\\   \hline 
  $L_1$ & always &$(\Sigma_{\phi},-3\Sigma_{\phi})$& $\{-3,0\}$& \text{Stable}\\ \hline
  $P_{3,6}$ & $\epsilon=\pm1$ & $ \left(\frac{\epsilon}{\sqrt{9 \mu +\epsilon}}, -\frac{3\epsilon}{\sqrt{9 \mu +\epsilon}}\right)$ & $\{-3,0\}$& \text{Stable} \\\hline
  $P_{4,7}$  & $\epsilon=\pm1$ & $\left(-\frac{\epsilon}{\sqrt{9 \mu +\epsilon}}, \frac{3\epsilon}{\sqrt{9 \mu +\epsilon}}\right)$ & $\{-3,0\}$& \text{Stable} \\\hline 
  $P_5$ & $\nexists$  &$(0,0)$& $\{-3,0\}$& \text{Stable}\\ \hline
    \end{tabular}
    \caption{Equilibrium points of system \eqref{syst1a} - \eqref{syst1b} for $\epsilon=+1$ with their eigenvalues and stability. $P_5$ which is a sink, but does not satisfy the condition \eqref{const1}}
    \label{tab:1}
\end{table}

\subsection{Analysis of the 2D flow}

In this section, we analyse the 2D flow associated with the dynamical system \eqref{syst1a}  and 
\eqref{syst1b}. We obtain the (lines of) equilibrium points of the system \eqref{syst1a}  and 
\eqref{syst1b} are summarised in Tab. \ref{tab:1}   for $\epsilon=\pm 1$  along with  with their coordinates, eigenvalues and stability. 

\subsubsection{Case $\epsilon=+1$}

In the case $\epsilon=+1$, there exists kinetic-dominated solutions, given by the points $P_1$ and $P_2$. They represent stiff solutions ($w_{\phi}=1$). 

There is a line of equilibrium points $L_1$ which corresponds to
\begin{equation}
   \dot{\phi} +  \frac{\sqrt{6}}{3} \omega_1^2 =0 \implies \phi = - \frac{\sqrt{6}}{3} \omega_1^2 t + c_1. 
\end{equation}
Then, from \eqref{FRW3}, we have at the lines of equilibrium points
\begin{align}
   & \dot{H}=  0 \implies H=H_0,  \implies a =  e^{H_0 (t-t_U)}. 
\end{align}
where $H_0$ satisfies $3 H_0^2 = \Lambda +\frac{ \omega_{1}^4 \epsilon }{3}$. 
That is a de Sitter solution.

Moreover, imposing the condition \eqref{const1}, the lines are reduced to the points  $P_3$
   and  $P_4$ that belongs to the lines of equilibrium points $3{\Sigma_\phi}+{\Sigma}=0$. 
   
For these equilibrium points, we have 
 $ \dot{\phi}= \sqrt{6} {\Sigma_\phi}_{1,2} H$ where $     {\Sigma_\phi}_{1,2}= \pm {1}/{\sqrt{9 \mu +1}}$.
 Hence,
 \begin{equation}
  \dot{H}+{\Sigma_\phi}_{1,2}  H  \left(3 {\Sigma_\phi}_{1,2}
   H + \sqrt{{\Lambda }/{(3\mu)}}\right)=0.
 \end{equation}
 That is, 
 \begin{small}
 \begin{equation}
     H  = \frac{\sqrt{3} H_0 \sqrt{\frac{\Lambda }{\mu }}}{\left(9 H_0 {\Sigma_\phi}_{1,2}+\sqrt{3}
   \sqrt{\frac{\Lambda }{\mu }}\right) e^{(t-t_U) {\Sigma_\phi}_{1,2}
    \sqrt{\frac{\Lambda }{3\mu }}}-9 H_0
   {\Sigma_\phi}_{1,2} }.
 \end{equation}
 \end{small}
The  line $L_1$ also contains the point  $P_5$, which is a sink but does not satisfy the condition \eqref{const1}.

\begin{figure*}[ht]
    \centering
    \includegraphics[scale=0.5]{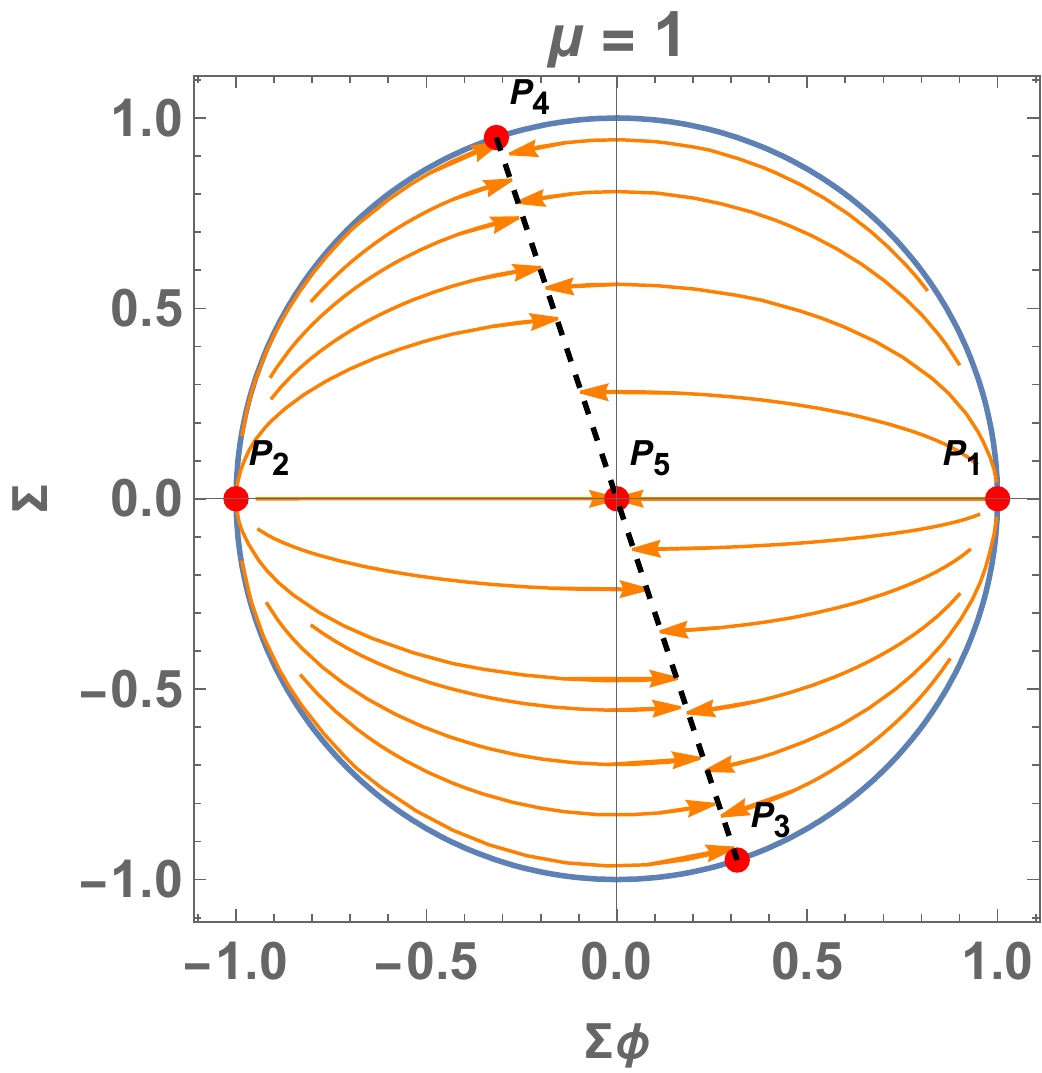}
\includegraphics[scale=0.5]{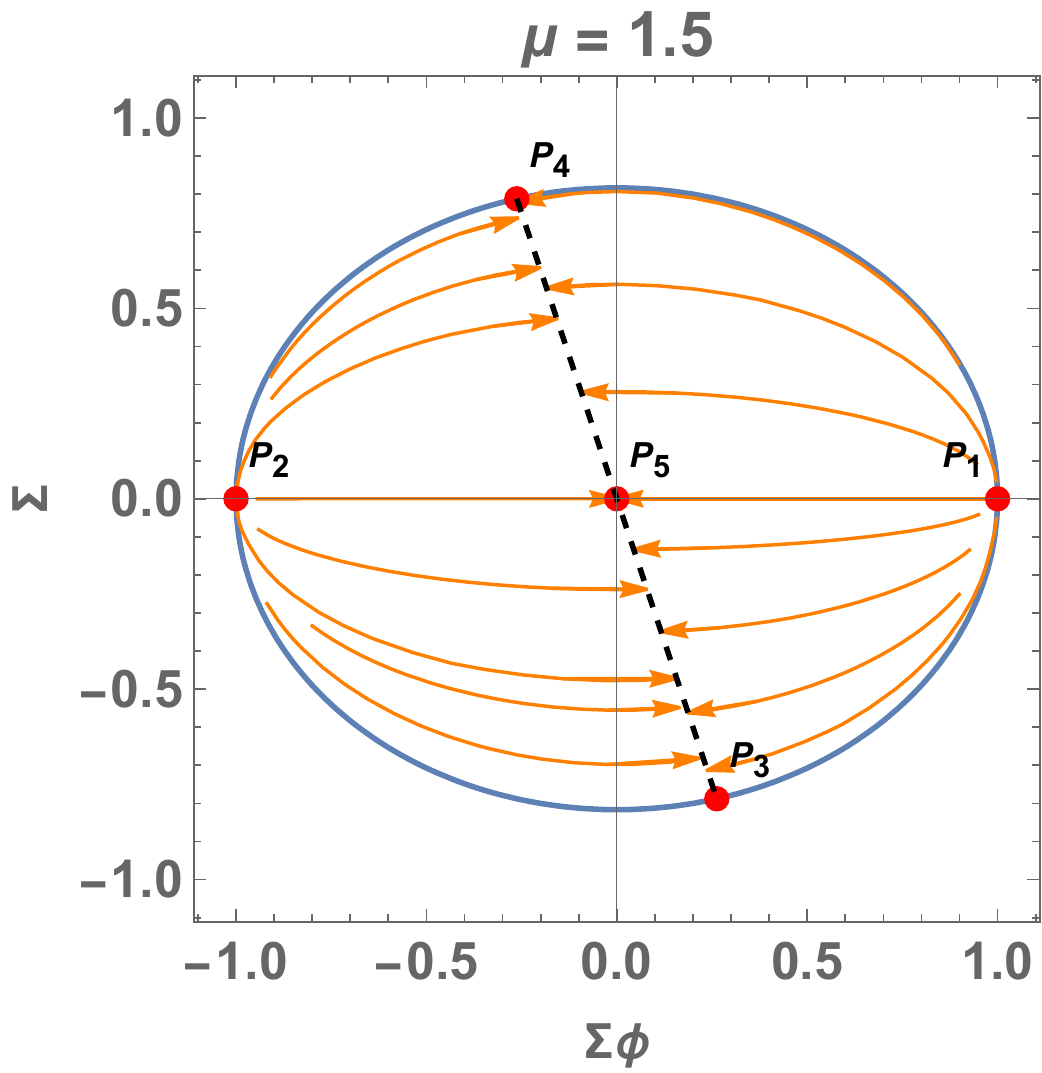}
\includegraphics[scale=0.5]{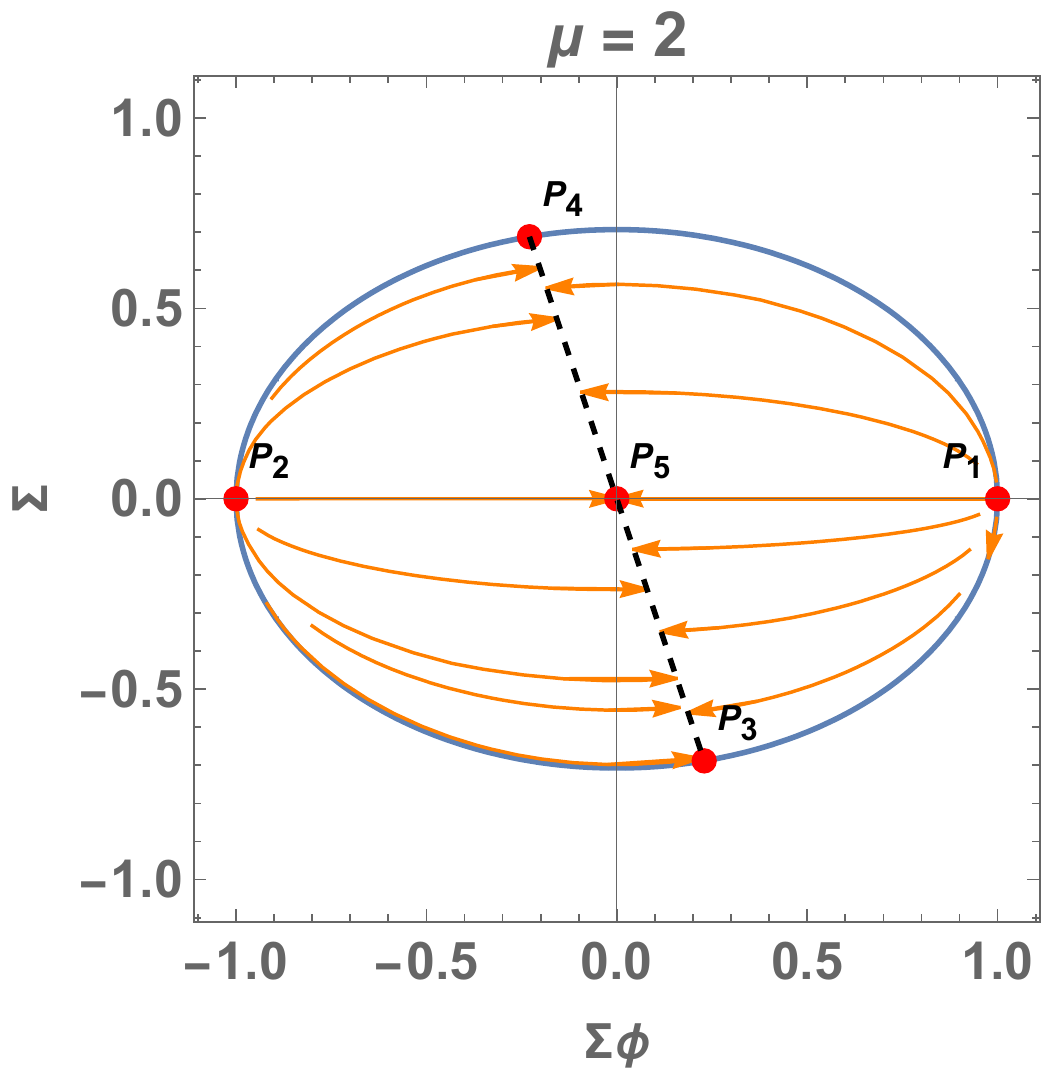}
    \caption{Phase Plot for system \eqref{syst1a}-\eqref{syst1b} for $\epsilon=+1$ and different values of $\mu$. The dashed black line corresponds to $L_1$.}
    \label{fig:1}
\end{figure*}

In Fig. \ref{fig:1} is presented a phase plot for system \eqref{syst1a}-\eqref{syst1b} for $\epsilon=+1$ and different values of $\mu$. The dashed black line corresponds to $L_1$.

\subsubsection{Case $\epsilon=-1$}

For the case $\epsilon=-1$, the equilibrium points of the system \eqref{syst1a}  and 
\eqref{syst1b} are 
as before the line $L_1$ which contains $P_5$ (which does not satisfies \eqref{const1}. Moreover, imposing the condition \eqref{const1}, the lines are reduced to the points  $P_6$
   and  $P_7$ that belongs to the lines of equilibrium points $3{\Sigma_\phi}+{\Sigma}=0$, where 
   ${\Sigma_\phi}_{1,2}= \mp {1}/{\sqrt{9 \mu -1}}$. 
   
   For these equilibrium points, we have 
 $ \dot{\phi}= \sqrt{6} {\Sigma_\phi}_{1,2} H$. Hence,
 \begin{equation}
  \dot{H} -{\Sigma_\phi}_{1,2}   H  \left(3 {\Sigma_\phi}_{1,2}
   H +\sqrt{ {\Lambda }/{(3\mu)}}\right)=0.
 \end{equation}
 That is, 
 \begin{small}
 \begin{equation}
     H  = \frac{\sqrt{3} H_0 \sqrt{\frac{\Lambda }{\mu }}}{\left(9 H_0 {\Sigma_\phi}_{1,2}+\sqrt{3}
   \sqrt{\frac{\Lambda }{\mu }}\right) e^{-(t-t_U) {\Sigma_\phi}_{1,2}
   \sqrt{\frac{\Lambda }{3\mu }}}-9 H_0
   {\Sigma_\phi}_{1,2} }.
 \end{equation} 
 \end{small}

\begin{figure*}[ht]
    \centering
    \includegraphics[scale=0.5]{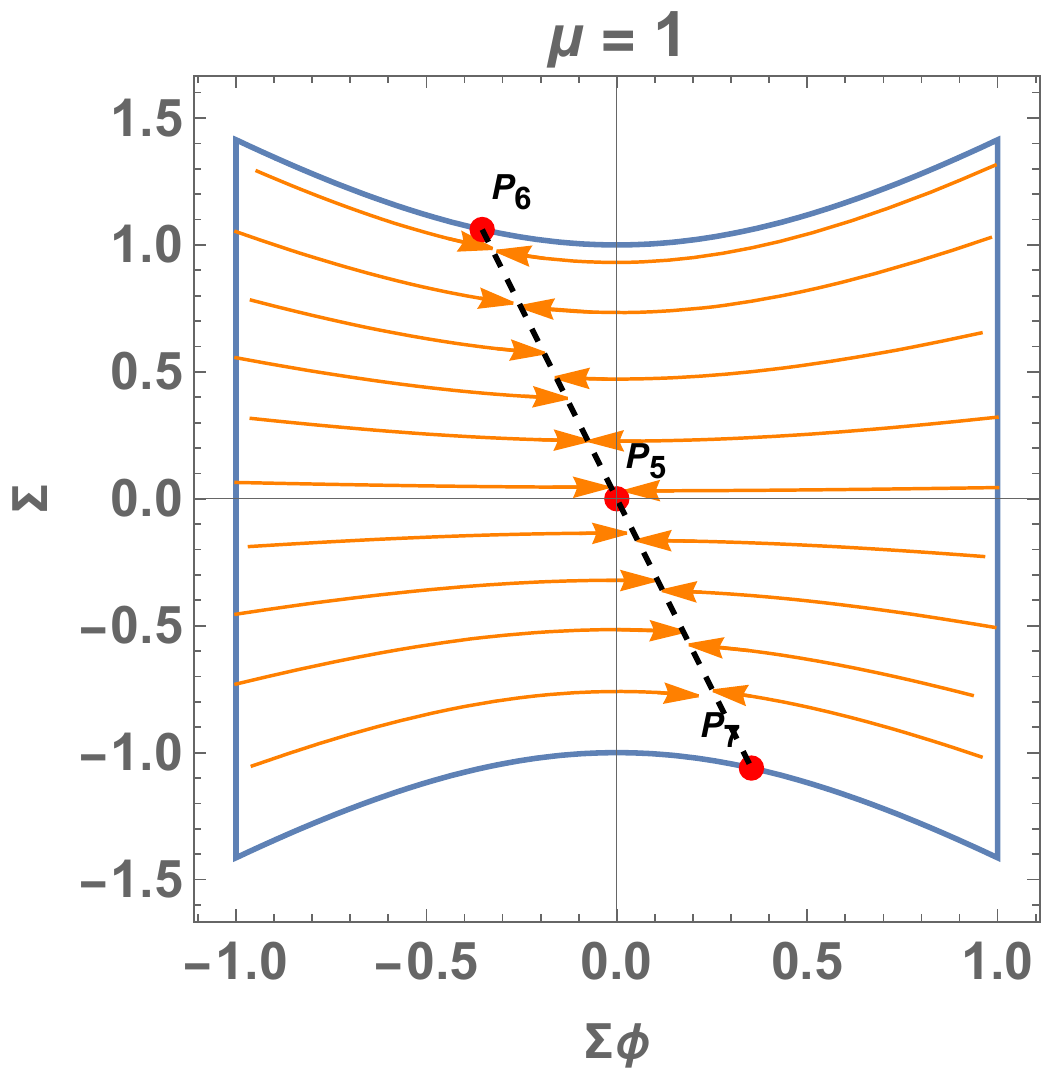}
\includegraphics[scale=0.5]{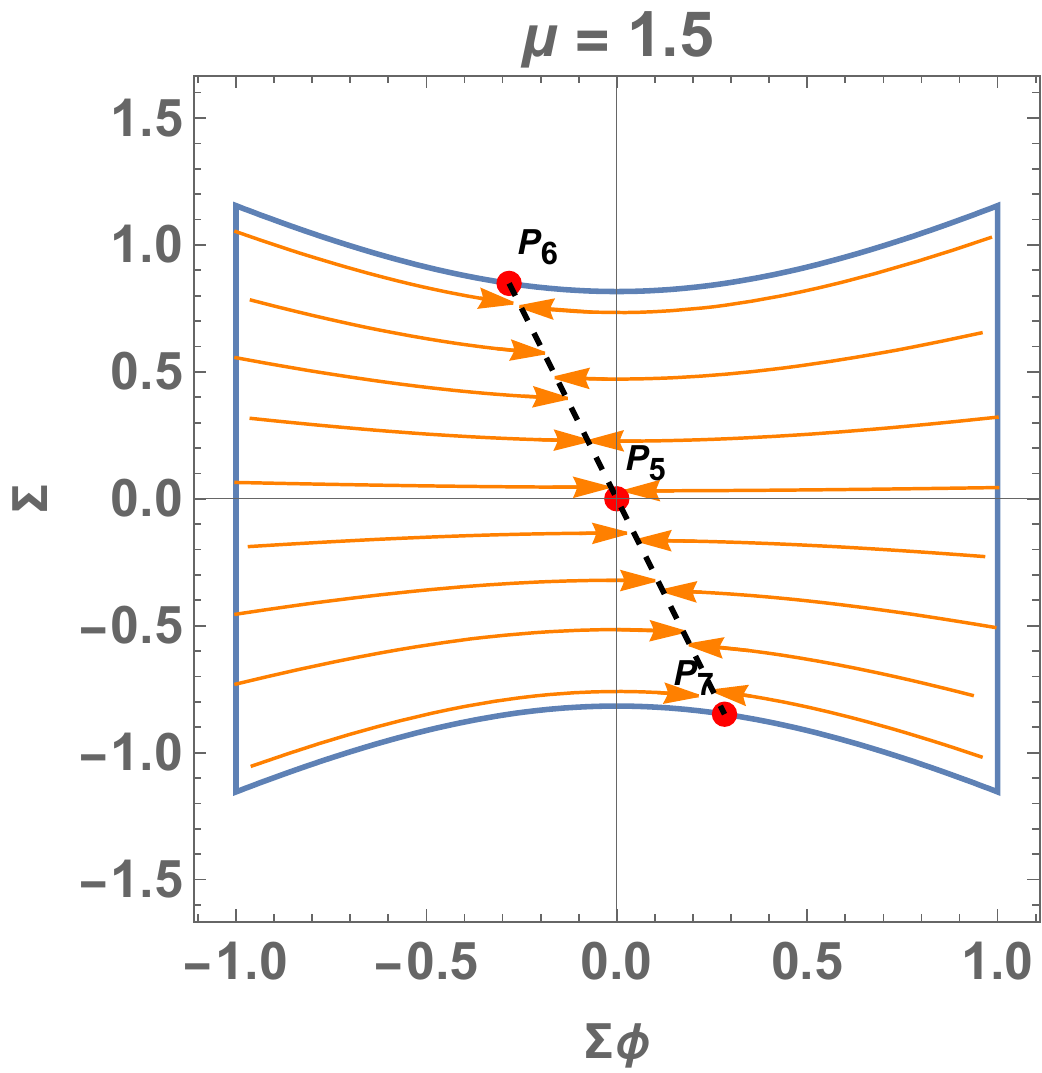}
\includegraphics[scale=0.5]{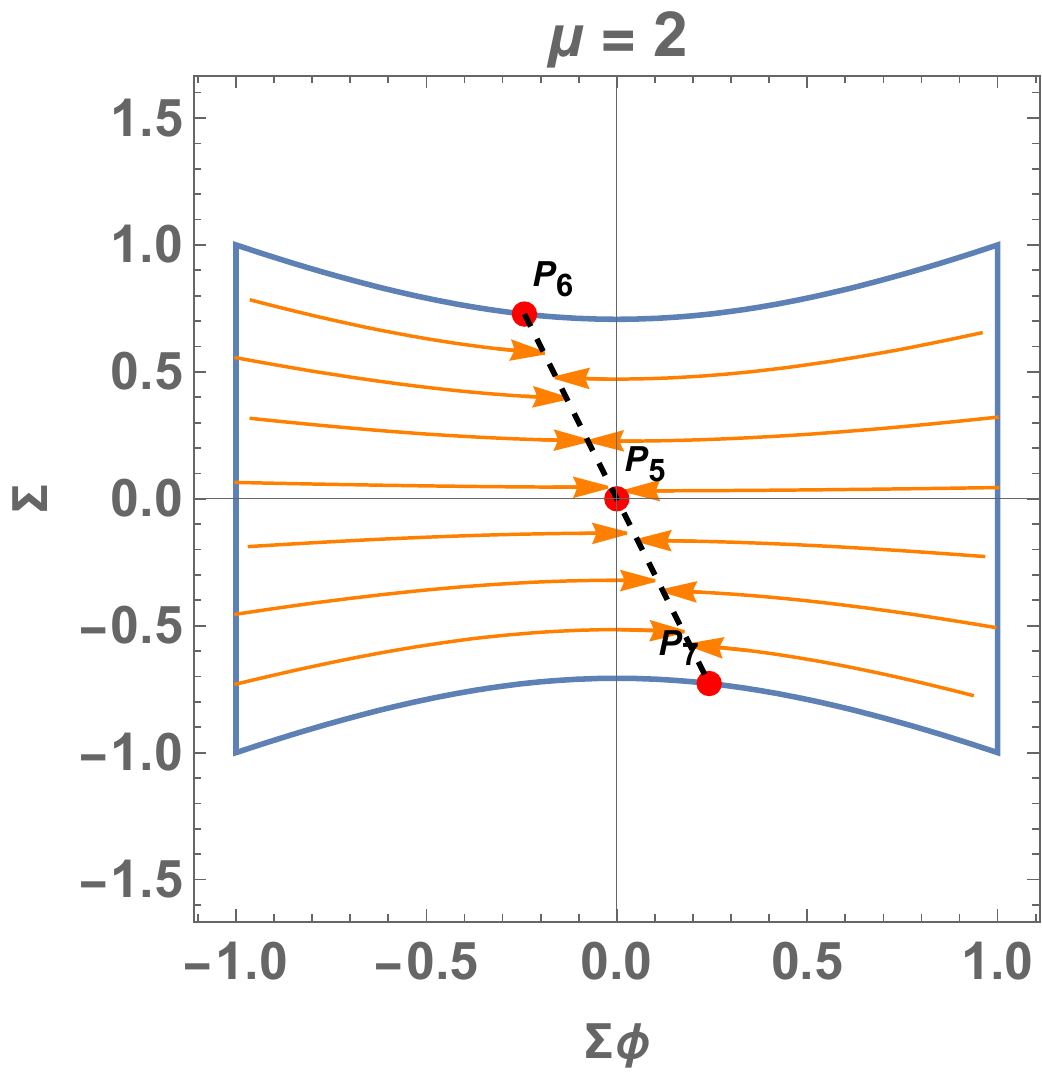}
    \caption{Phase plot for system \eqref{syst1a}  - 
\eqref{syst1b} for $\epsilon=-1$ and different values of $\mu$. The dashed black line corresponds to $L_1$.}
    \label{fig:1b}
\end{figure*}

In Fig. \ref{fig:1b} is presented a phase plot for system \eqref{syst1a}-\eqref{syst1b} for $\epsilon=-1$ and different values of $\mu$. The dashed black line corresponds to $L_1$.

\subsection{1D reduced system}

Using \eqref{const1} to reduce the dimensionality and for ${\Sigma}\geq0 $ we have  \begin{equation} {\Sigma}=\sqrt{(1-
     \Sigma_\phi^2 \epsilon)/\mu}.
\end{equation}
Then, we have the reduced dynamical system
\begin{align}
 & \Sigma_\phi^{\prime}= -  \left(3 \Sigma_\phi+\sqrt{(1-
     \Sigma_\phi^2 \epsilon)/\mu}\right) \left(1-\Sigma_\phi^2 \epsilon \right), \label{reduced1a}\
 \end{align}
This patch cover only the equilibrium points \eqref{syst1a}- \eqref{syst1b} with $\Sigma>0$ and $\Sigma_\phi=0$, say,  $P_1: \Sigma_\phi= 1$, 
$ P_2:  \Sigma_\phi= -1$  and $P_4: \Sigma_\phi= -1/{\sqrt{9 \mu +1}}$. 
     Moreover, the equilibrium point of the 1D system \eqref{reduced1a} is $P_6: {\Sigma_\phi}= - {1}/{\sqrt{9
   \mu -1}}$. 
 
 \begin{table}[h]
    \centering
    \begin{tabular}{|c|c|c|c|c|c|}
    \hline
  \text{Label } & \text{Existence} & \text{Coordinates}\; $\Sigma_{\phi}$ & \text{Eigenvalue}& \text{Stability}\\\hline   
  $P_{1,2}$ & $\epsilon=+1$&$\pm 1$& 6 & \text{Unstable}\\   \hline 
  $P_{3}$ & $\epsilon=+1, \mu>-\frac{1}{9}$&$\frac{1}{\sqrt{9 \mu +1}}$& $-3$& \text{Stable}\\ \hline
   $P_{6}$ & $\epsilon=-1, \mu>\frac{1}{9}$&$-\frac{1}{\sqrt{9
   \mu -1}}$& $-3$& \text{Stable}\\ \hline
    \end{tabular}
    \caption{Equilibrium points of system  \eqref{reduced1a} for $\epsilon=\pm 1$ with their eigenvalues and stability}
    \label{tab:1 1D}
\end{table} 
Tab. \ref{tab:1 1D} presents the equilibrium points of system  \eqref{reduced1a} for $\epsilon=\pm 1$ with their eigenvalues and stability. 

\begin{figure*}[ht]
    \centering
    \includegraphics[scale=0.5]{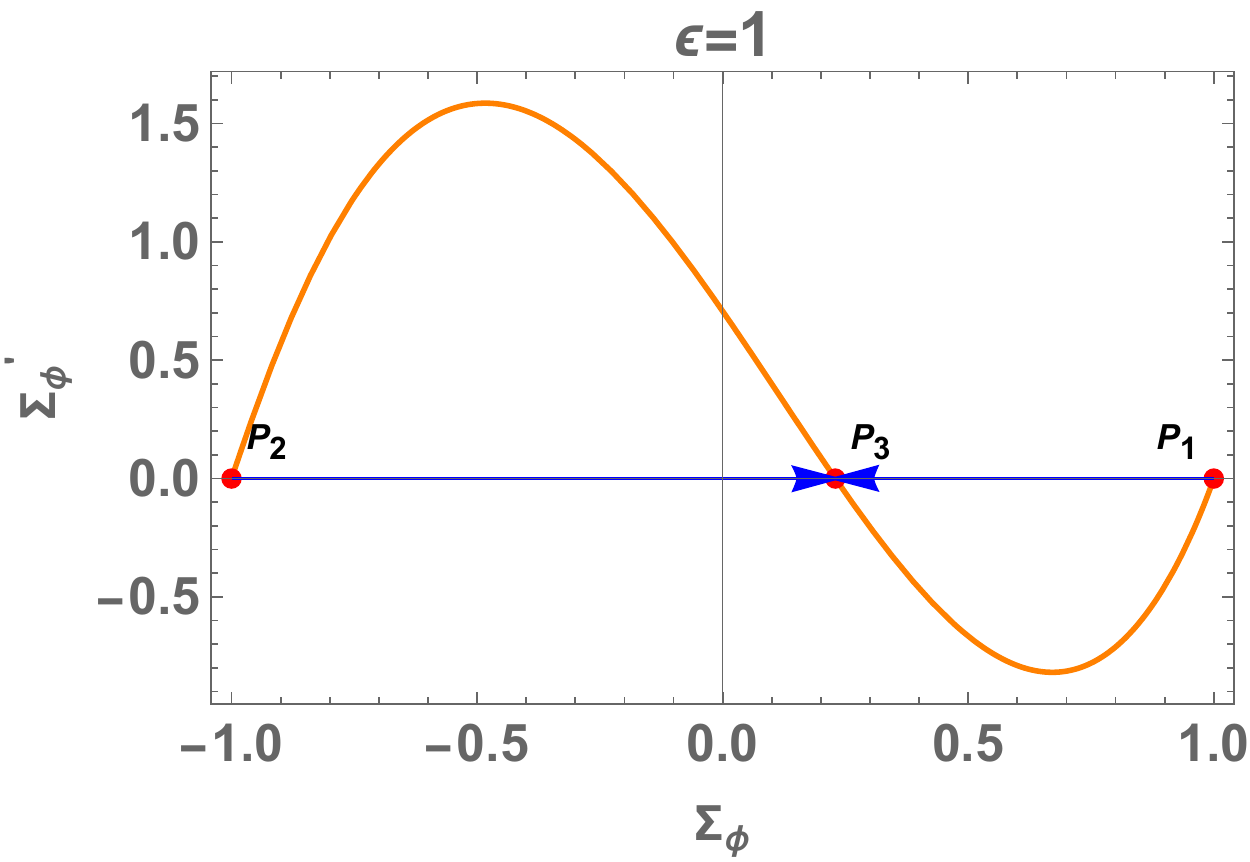}
     \includegraphics[scale=0.5]{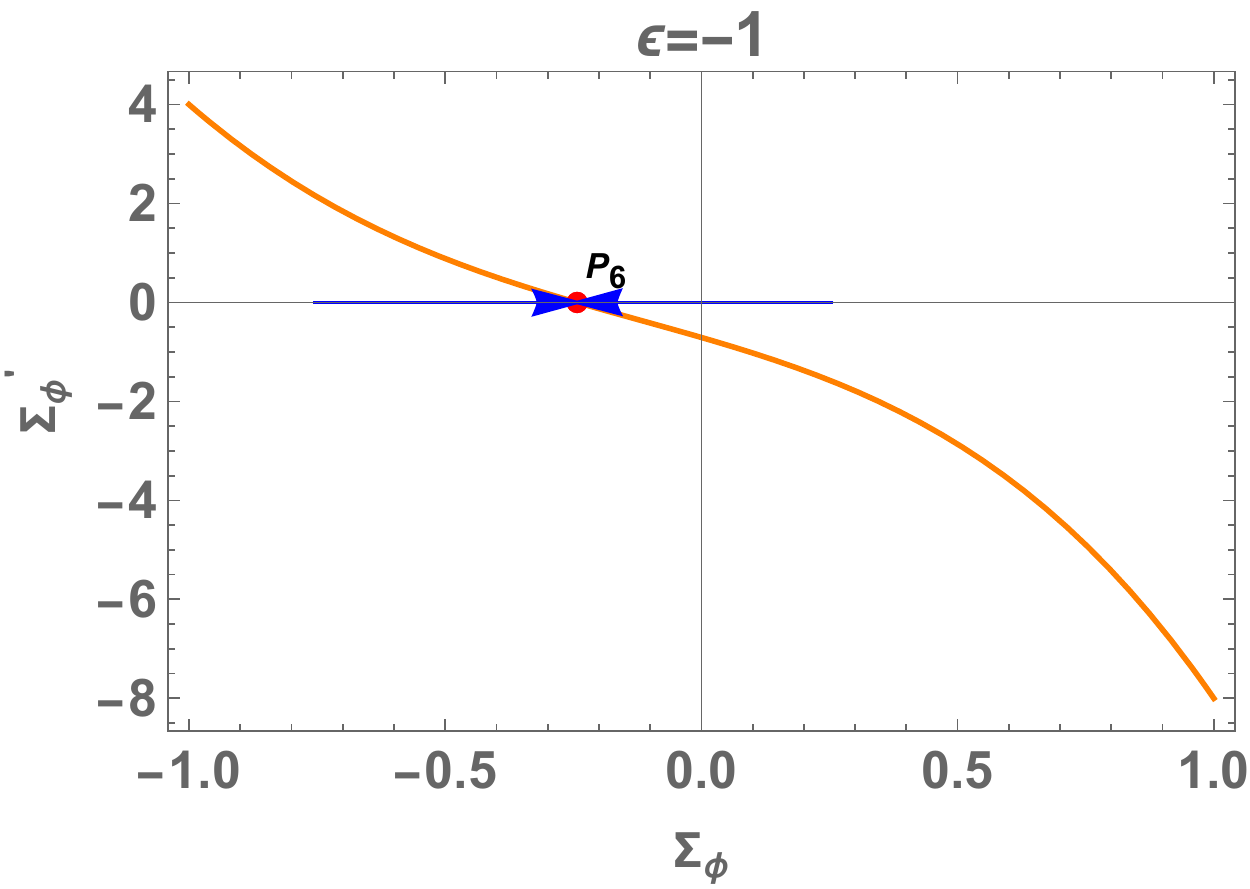}
    \caption{Phase plot for the reduced 1D equation \eqref{reduced1a} for $\epsilon=\pm 1$ and $\mu=2$.}
    \label{fig:2}
\end{figure*}

In Fig. \ref{fig:2} is presented a phase plot for the reduced 1D equation \eqref{reduced1a} for $\epsilon=\pm 1$ and $\mu=2$. 
 
 On the other hand, using 
 \begin{equation} {\Sigma}= -\sqrt{(1-
     \Sigma_\phi^2 \epsilon)/\mu},
\end{equation}
we have the reduced dynamical system
\begin{align}
 & \Sigma_\phi^{\prime}= -  \left(3 \Sigma_\phi- {\sqrt{(1-
     \Sigma_\phi^2 \epsilon)/\mu}}\right) \left(1-\Sigma_\phi^2 \epsilon \right), \label{reduced1b}\
 \end{align}
This patch cover only the equilibrium points \eqref{syst1a}- \eqref{syst1b} with $\Sigma<0$  and $\Sigma_\phi=0$, say,  $P_1: \Sigma_\phi= 1$ , $P_2: \Sigma_\phi= -1$
   and $P_3: {\Sigma_\phi}:=  {1}/{\sqrt{9 \mu +1}}$. Moreover, the equilibrium point of the 1D system \eqref{reduced1b} is 
$P_7: {\Sigma_\phi}=  {1}/{\sqrt{9 \mu -1}}$, 
that belongs to the lines of equilibrium points $3{\Sigma_\phi}+{\Sigma}=0$.

\begin{table}[h]
    \centering
    \begin{tabular}{|c|c|c|c|c|c|}
    \hline
  \text{Label} & \text{Existence} & \text{Coordinates}\; $\Sigma_{\phi}$ & \text{Eigenvalue}& \text{Stability}\\\hline   
  $P_{1,2}$ & $\pm 1$ & $\epsilon=+1$& 6 & \text{Unstable}\\  \hline 
  $P_{4}$ & $\epsilon=+1, \mu>-\frac{1}{9}$ &$-\frac{1}{\sqrt{9 \mu +1}}$& $-3$& \text{Stable}\\ \hline
  $P_{7}$ & $\epsilon=-1, \mu>\frac{1}{9}$&$\frac{1}{\sqrt{9 \mu -1}}$& $-3$& \text{Stable}\\ \hline
    \end{tabular}
    \caption{Equilibrium points of system \eqref{reduced1b} for $\epsilon=\pm 1$ with their eigenvalues and stability}
    \label{tab:2 1D}
\end{table}
In Tab. \ref{tab:2 1D} are presented the equilibrium points of system \eqref{reduced1b} for $\epsilon=\pm 1$ with their eigenvalues and stability. 
\begin{figure*}[ht]
    \centering
     \includegraphics[scale=0.5]{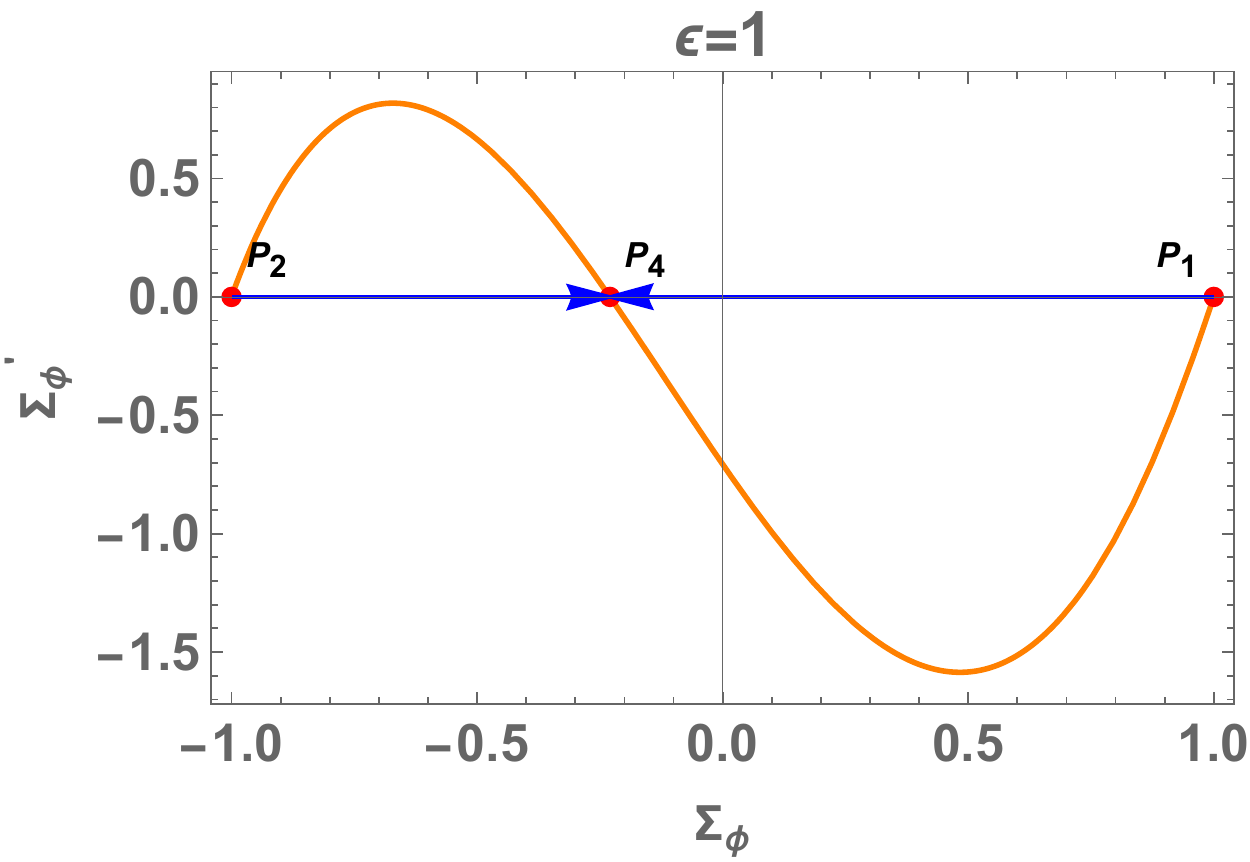}
     \includegraphics[scale=0.5]{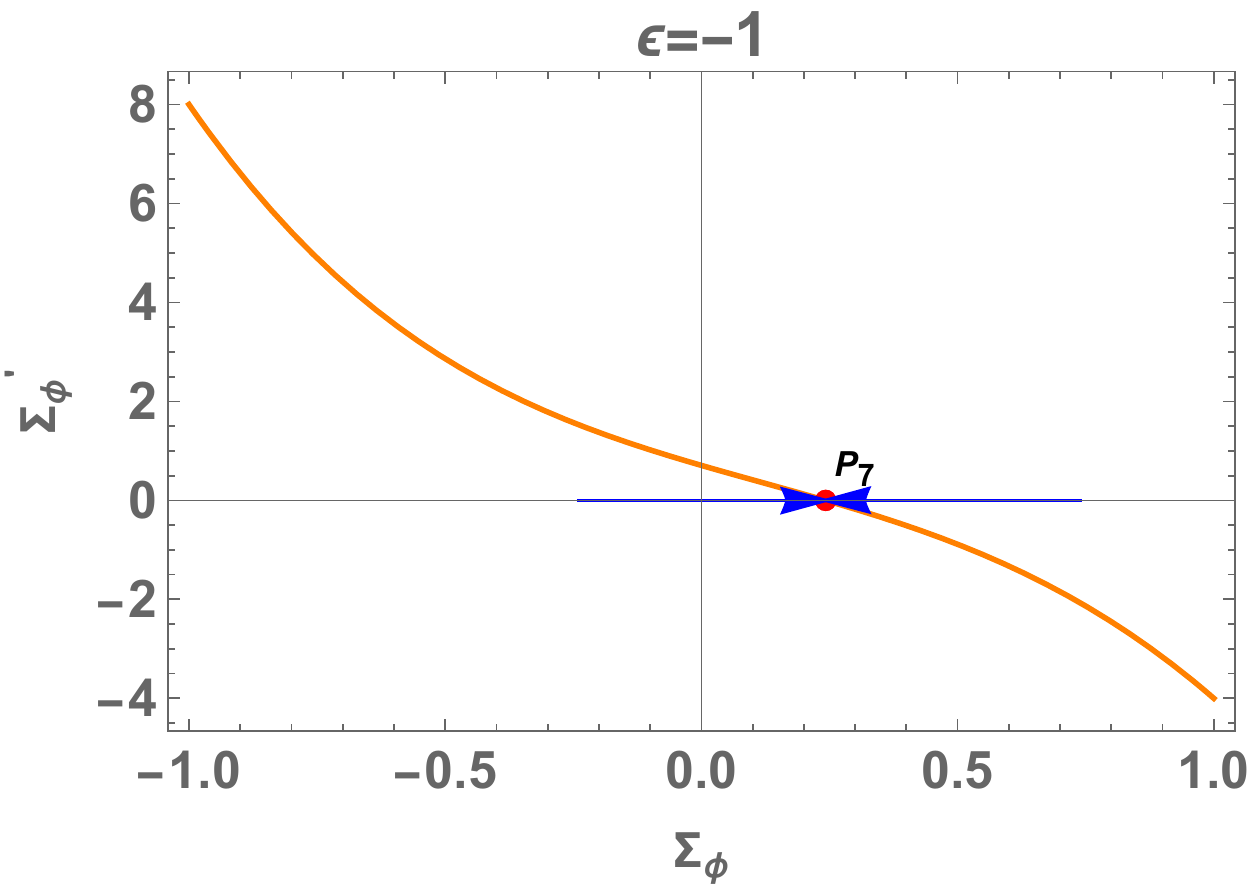}
    \caption{Phase plot for the reduced 1D equation \eqref{reduced1b} for $\epsilon=\pm 1$ and $\mu=2$.}
    \label{fig:2b}
\end{figure*}
Fig. \ref{fig:2b} presents a phase plot for the reduced 1D equation \eqref{reduced1b} for $\epsilon=\pm 1$ and $\mu=2$.
 
\section{Dynamical systems analysis by including matter}

We define 
\begin{equation}
    {\Sigma_\phi}=\frac{\dot{\phi}}{\sqrt{6}H}, \; {\Sigma}=  \frac{\omega_1^2}{H}, \; \Omega_m=\frac{\rho_m}{3 H^2},
\end{equation}
which satisfy \begin{equation}
     {\Sigma_\phi}^2 \epsilon +  \mu {\Sigma}^2 + \Omega_m=1. \label{cons_2}
\end{equation}

\subsection{3D system}

The system \eqref{FRW1}, \eqref{FRW2} and\eqref{FRW1} can be written as the dynamical system given by 
\begin{small}
\begin{align}
& {\Sigma_\phi}^{\prime}=   \frac{3}{2} (w_m+1) {\Sigma_\phi} \Omega_m+(3 {\Sigma_\phi}+{\Sigma})
   \left({\Sigma_\phi}^2 \epsilon -1\right),  \label{syst2a}\\
& {\Sigma}^{\prime}=\frac{3}{2} (w_m+1) {\Sigma}
   \Omega_m+{\Sigma_\phi} {\Sigma} \epsilon  (3 {\Sigma_\phi}+{\Sigma}), \label{syst2b}\\
& \Omega_m^{\prime}=\Omega_m (3
   w_m (\Omega_m-1)+2 {\Sigma_\phi} \epsilon  (3 {\Sigma_\phi}+{\Sigma})+3
   (\Omega_m-1)). \label{syst2c}
\end{align}
\end{small}

\begin{table*}[!t]
    \centering
    \begin{tabular}{|c|c|c|c|c|}
    \hline
  \text{Label}& \text{Existence} & \text{Coordinates}\; $(\Sigma_{\phi},  \Sigma,\Omega_m)$ & \text{Eigenvalues }& \text{Stability}\\\hline 
   $P_{1,2}$& $\epsilon= + 1$&$(\pm 1,0,0)$& $\{3, 6,3(1-\omega_m)\}$ & \text{Unstable}\\ \hline
   $L_1$&always &$(\Sigma_{\phi},-3\Sigma_{\phi},0)$& $\{-3,0,-3 (\omega_m +1)\}$& \text{Stable}\\ \hline
   $P_{3,6}$& $\epsilon= \pm 1$ & $ \left(\frac{\epsilon}{\sqrt{9 \mu +\epsilon}}, -\frac{3\epsilon}{\sqrt{9 \mu +\epsilon}}, 0\right)$ & $\{-3,0, -3 (\omega_m +1)\}$& \text{Stable} \\\hline
  $P_{4,7}$& $\epsilon= \pm 1$& $\left(-\frac{\epsilon}{\sqrt{9 \mu +\epsilon}}, \frac{3\epsilon}{\sqrt{9 \mu +\epsilon}}, 0\right)$ & $\{-3,0, -3 (\omega_m +1)\}$& \text{Stable} \\\hline 
  $P_{5}$& $\nexists$ &$( 0,0,0)$& $\{-3,0,-3 (\omega_m +1)\}$ & \text{Stable} \\ \hline 
  $M$& $\epsilon= \pm 1$ &$(0,0,1)$& $\left\{\frac{3 (\omega_m -1)}{2},\frac{3 (\omega_m +1)}{2},3 (\omega_m +1)\right\}$ & \text{Stable for } $\omega_m=-1$\\&&&&\text{Unstable for }$\omega_m=1$\\&&&&\text{Saddle otherwise}\\ \hline
    \end{tabular}
    \caption{Equilibrium points of system \eqref{syst2a}, \eqref{syst2b} and \eqref{syst2c} for $\epsilon=\pm 1$ with their eigenvalues and stability}
    \label{tab:1A}
\end{table*}

In Tab. \ref{tab:1A} are presented the equilibrium points of system \eqref{syst2a}, \eqref{syst2b} and \eqref{syst2c} for $\epsilon=\pm 1$ with their eigenvalues and stability.

\subsubsection{Case $\epsilon=+1$}

For the case $\epsilon=+1$, the equilibrium points of the system \eqref{syst2a},
\eqref{syst2b}, and 
\eqref{syst2c} are  $P_1$, $P_2$ which are kinetic dominated solutions. The line of equilibrium points $L_1$ which represents de Sitter solutions. This line contains the points $P_3$, $P_4$ and $P_5$. Additionally, we have the matter-dominated solution $M:\left({\Sigma_\phi}, {\Sigma}, \Omega_m\right)=(0, 0, 1)$.

\begin{figure*}[ht]
    \centering
     \includegraphics[scale=0.85]{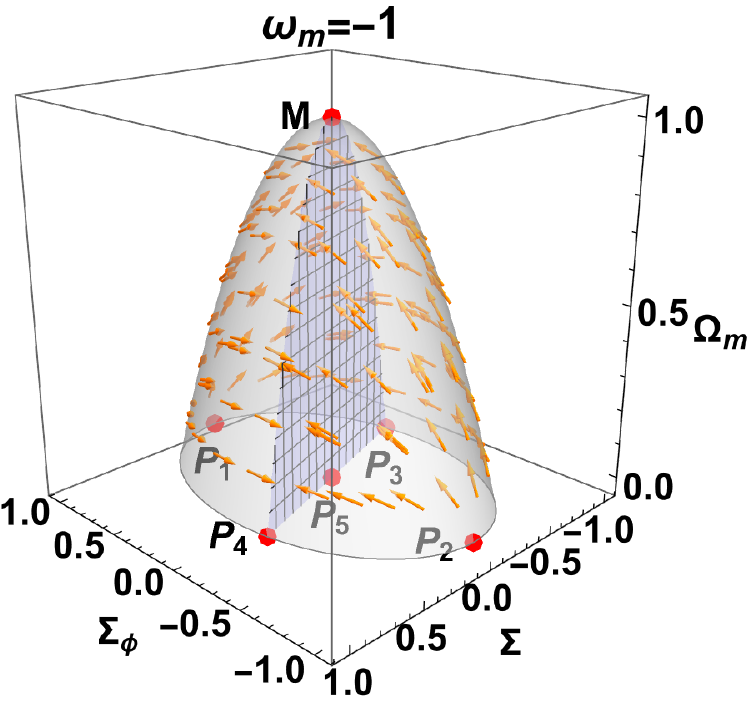}
     \includegraphics[scale=0.85]{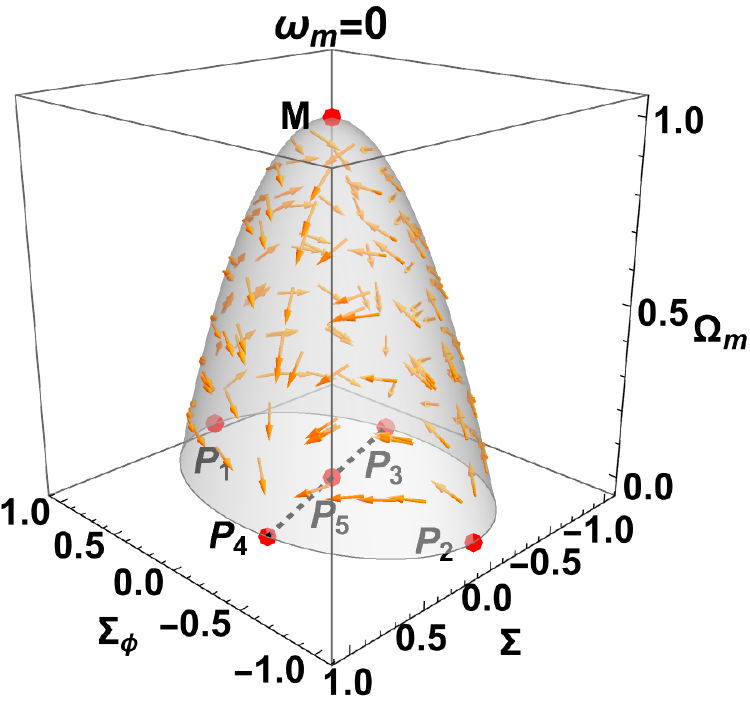}
     \includegraphics[scale=0.85]{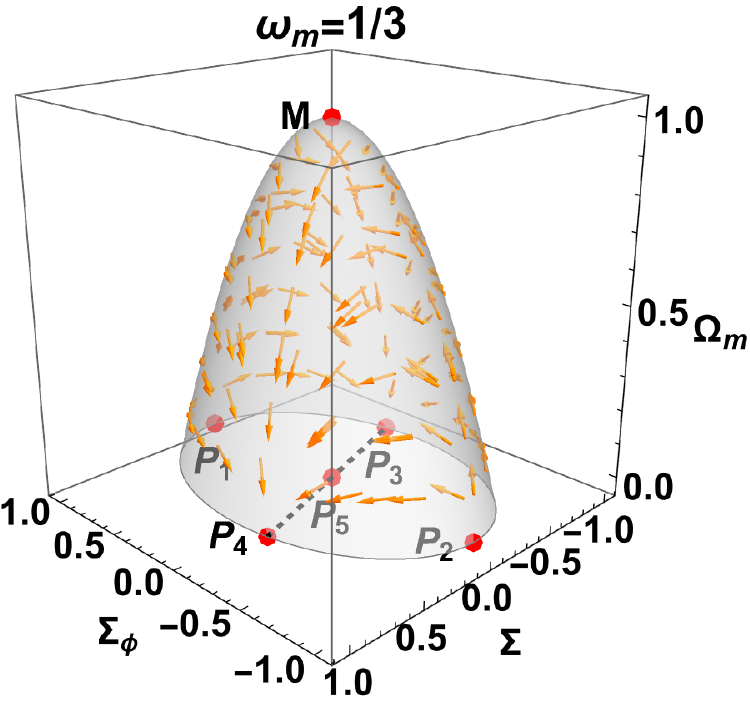}
     \includegraphics[scale=0.85]{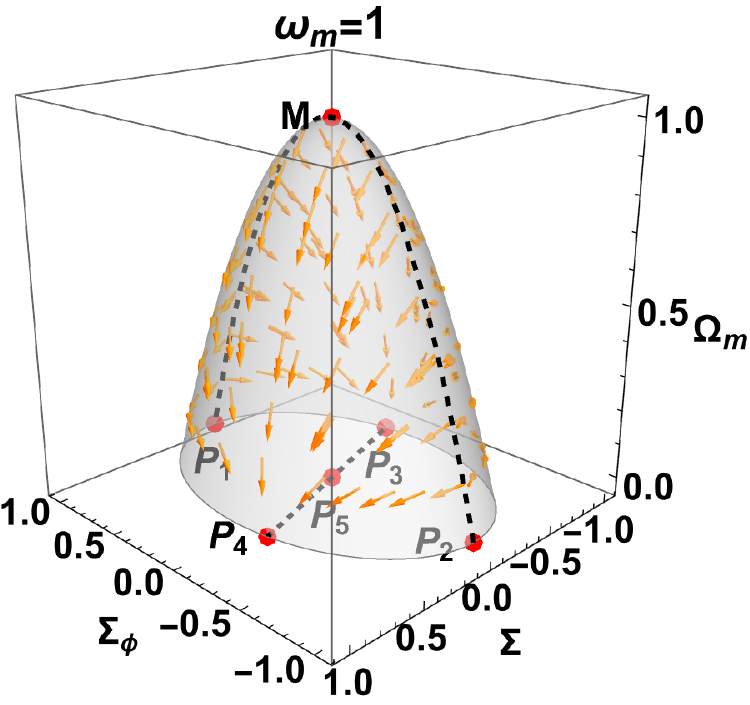}
    \caption{3D phase plot for system \eqref{syst2a}, \eqref{syst2b} and \eqref{syst2c} for $\epsilon=+1$, $\mu=2$ and different values of $\omega_m$.}
    \label{fig:3}
\end{figure*}

In  Fig. \ref{fig:3} is presented a 3D phase plot for system \eqref{syst2a}, \eqref{syst2b} and \eqref{syst2c} for $\epsilon=1$, $\mu=2$ and different values of $\omega_m$. 

\subsubsection{Case $\epsilon=-1$}

For the case $\epsilon=-1$, the equilibrium points of the system \eqref{syst2a},
\eqref{syst2b}, and 
\eqref{syst2c} are  the line of equilibrium points $L_1$ which represents de Sitter solutions. This line contains the points $P_5$, $P_6$ and $P_7$. Additionally, we have the matter-dominated solution $M:\left({\Sigma_\phi}, {\Sigma}, \Omega_m\right)=(0, 0, 1)$.

\begin{figure*}[ht]
    \centering
     \includegraphics[scale=0.85]{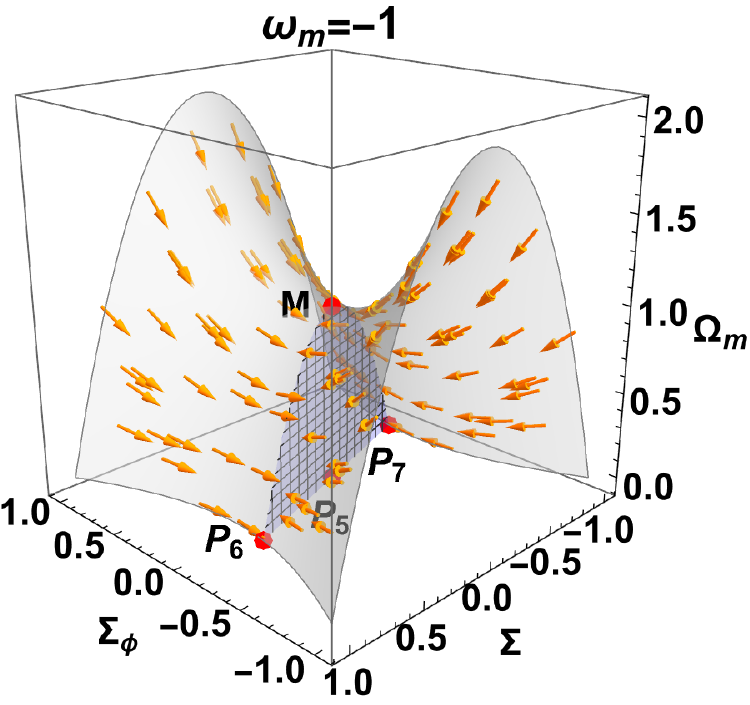}
     \includegraphics[scale=0.85]{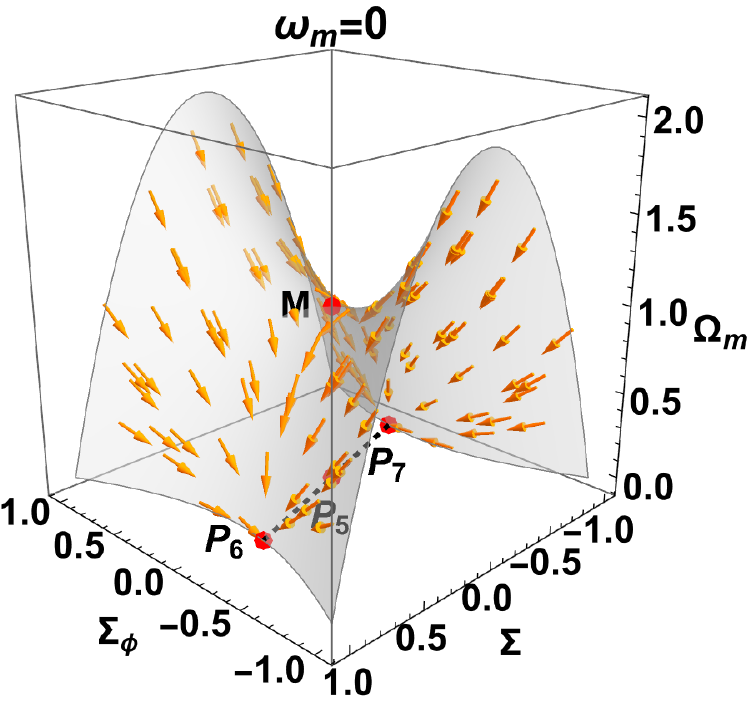}
     \includegraphics[scale=0.85]{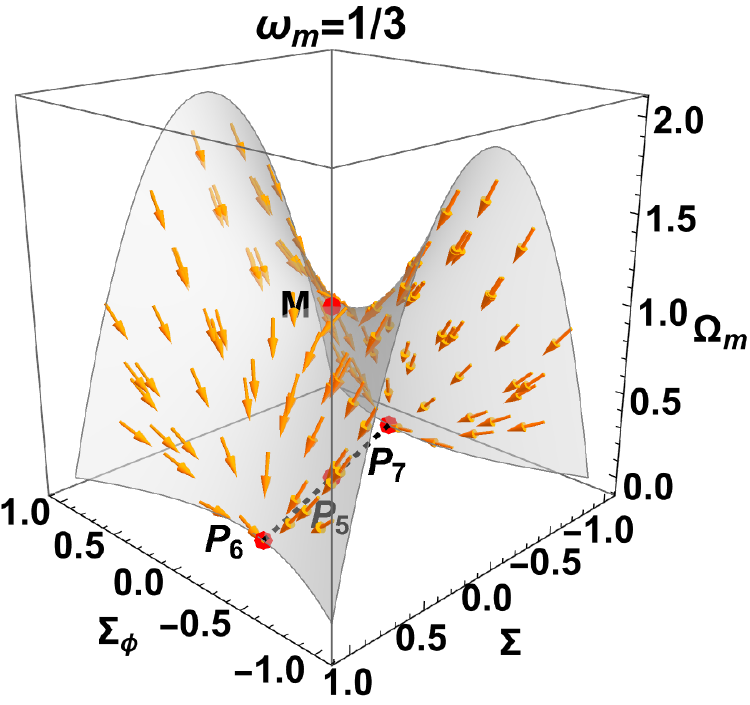}
     \includegraphics[scale=0.85]{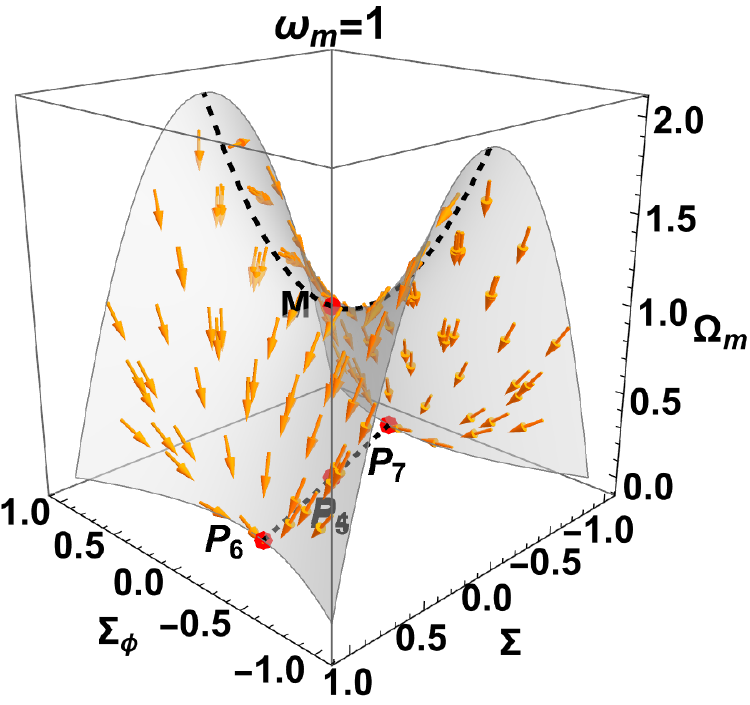}
    \caption{3D phase plot for system \eqref{syst2a}, \eqref{syst2b} and \eqref{syst2c} for $\epsilon=-1$, $\mu=2$ and different values of $\omega_m$.}
    \label{fig:5}
\end{figure*}

In Fig. \ref{fig:5} is presented a 3D phase plot for system \eqref{syst2a}, \eqref{syst2b} and \eqref{syst2c} for $\epsilon=-1$, $\mu=2$ and different values of $\omega_m$.
 
\subsection{Reduced 2D system}

Eliminating $\Omega_m$ from \eqref{cons_2} to obtain the reduced system\
\begin{small}
\begin{align}
  & {\Sigma_\phi}^{\prime}=-\frac{1}{2} \left({\Sigma_\phi}^2 \epsilon -1\right) (3 (w_m-1)
   {\Sigma_\phi}-2 {\Sigma})-\frac{3}{2} \mu  (w_m+1) {\Sigma_\phi} {\Sigma}^2, \label{reduced2a}\\
   & {\Sigma}^{\prime}=\frac{3}{2}
   (w_m+1) {\Sigma} \left(1-{\Sigma_\phi}^2  \epsilon -\mu  {\Sigma}^2\right)+{\Sigma_\phi} {\Sigma}
   \epsilon  (3 {\Sigma_\phi}+{\Sigma}). \label{reduced2b}
\end{align}
\end{small}

\subsubsection{Case $\epsilon=+1$}

The equilibrium points of the system \eqref{reduced2a} and \eqref{reduced2b} are $P_{1,2}$
$L_1$, $P_3$,  $P_4$  and $M$, summarised in Tab. \ref{tab:A1}. 
   
   \begin{table*}[!t]
    \centering
    \begin{tabular}{|c|c|c|c|}
    \hline
  \text{Label}&  \text{Coordinates}\; $(\Sigma_{\phi},  \Sigma)$ & \text{Eigenvalues }& \text{Stability}\\\hline 
  $P_{1,2}$&$(\pm 1,0)$& $\{3, 6\}$ & \text{Unstable}\\ \hline
  $L_1$&$(\Sigma_{\phi},-3\Sigma_{\phi})$& $\{-3, -3 (\omega +1)\}$& \text{Stable}\\ \hline
  $P_3$ & $ \left(\frac{1}{\sqrt{9 \mu +1}}, -\frac{3}{\sqrt{9 \mu +1}} \right)$ & $\{-3,  -3 (\omega +1)\}$& \text{Stable} \\\hline
  $P_4$ & $\left(-\frac{1}{\sqrt{9 \mu +1}}, \frac{3}{\sqrt{9 \mu +1}} \right)$ & $\{-3,  -3 (\omega +1)\}$& \text{Stable} \\\hline 
  $M$&$(0,0)$& $\left\{\frac{3 (\omega_m -1)}{2}, 3 (\omega_m +1)\right\}$ & \text{Stable for } $\omega=-1$\\&&&\text{Unstable for }$\omega=1$\\&&&\text{Saddle otherwise}\\ \hline
    \end{tabular}
    \caption{Equilibrium points of system \eqref{reduced2a} and \eqref{reduced2b} for $\epsilon=+1$ with their eigenvalues and stability}
    \label{tab:A1}
\end{table*}

\begin{figure*}[ht]
    \centering
     \includegraphics[scale=0.7]{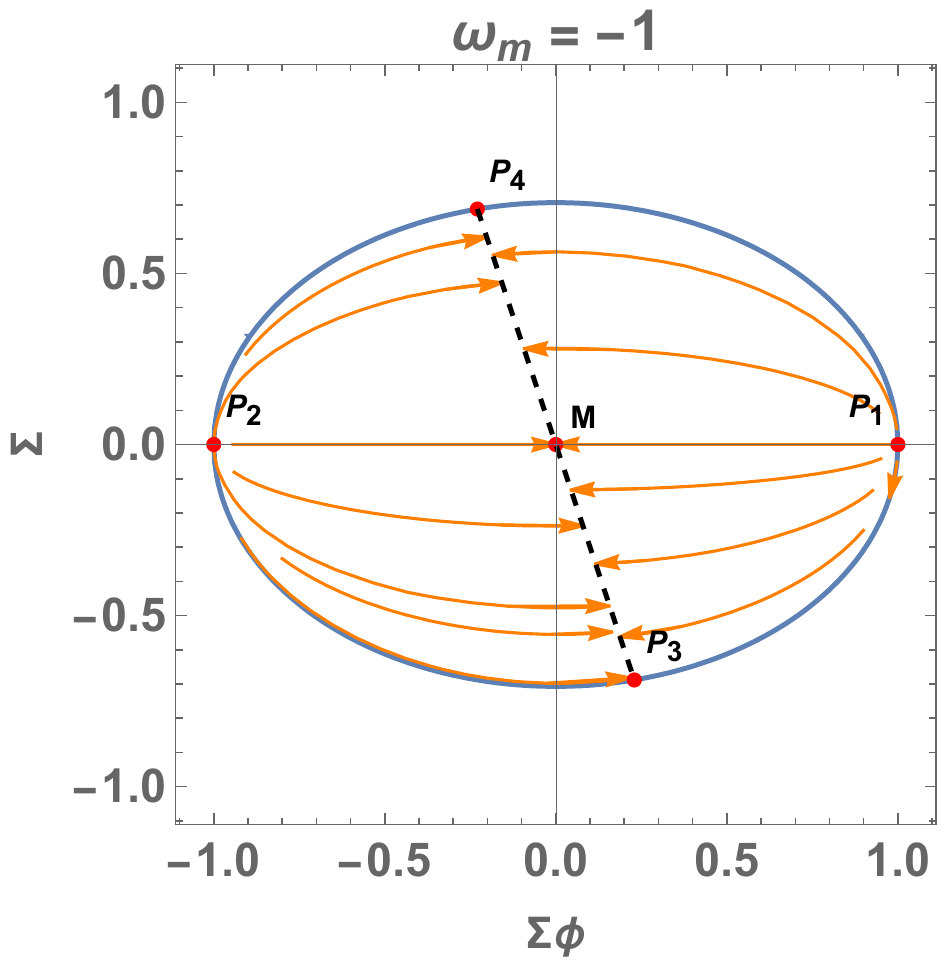}
     \includegraphics[scale=0.7]{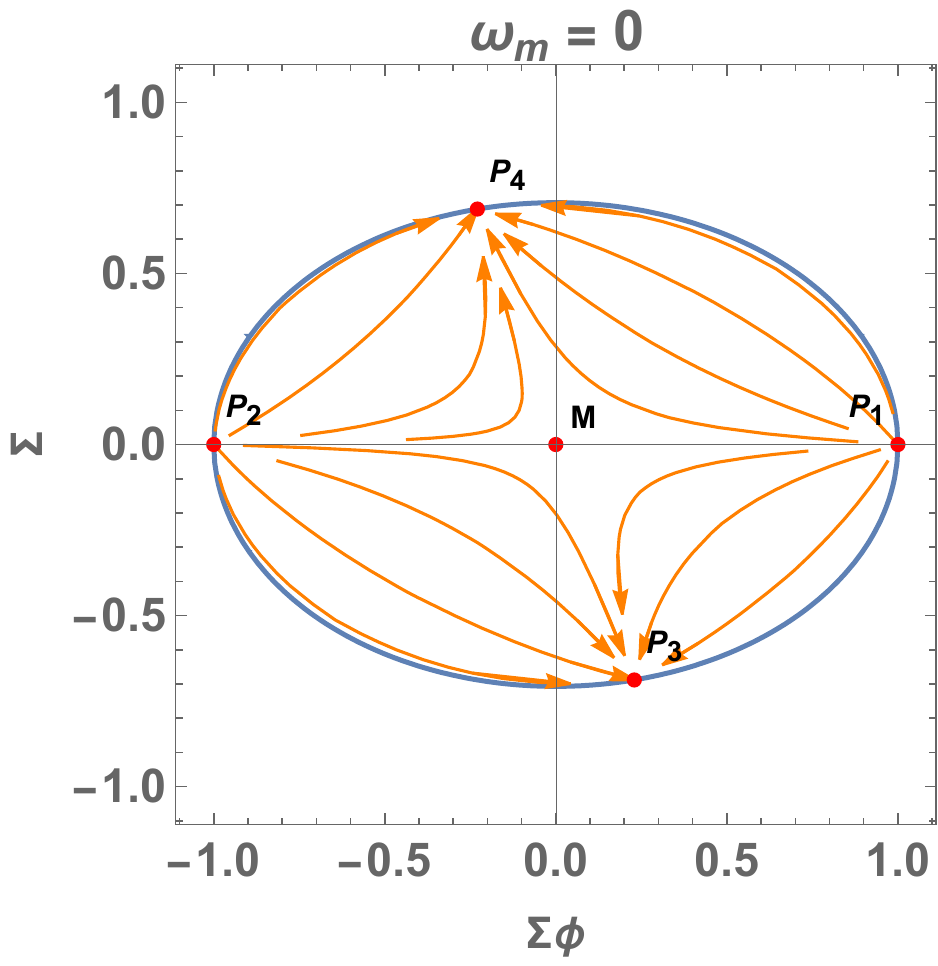}
     \includegraphics[scale=0.7]{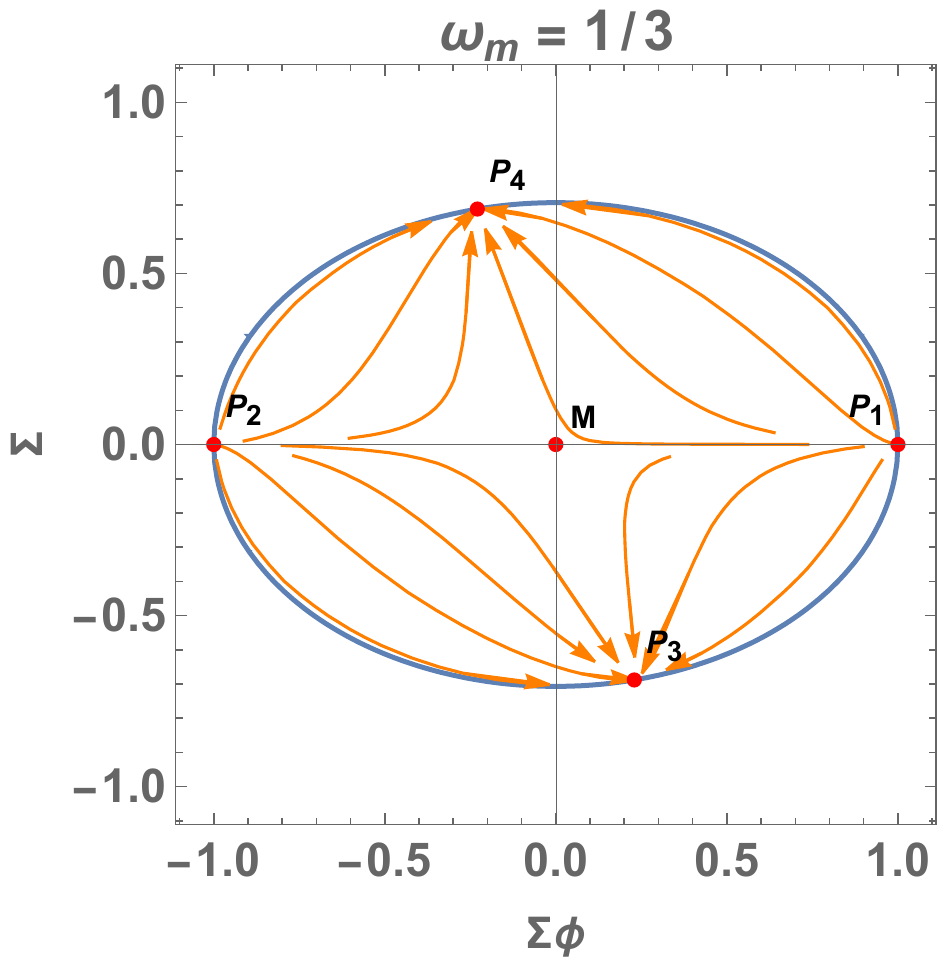}
     \includegraphics[scale=0.7]{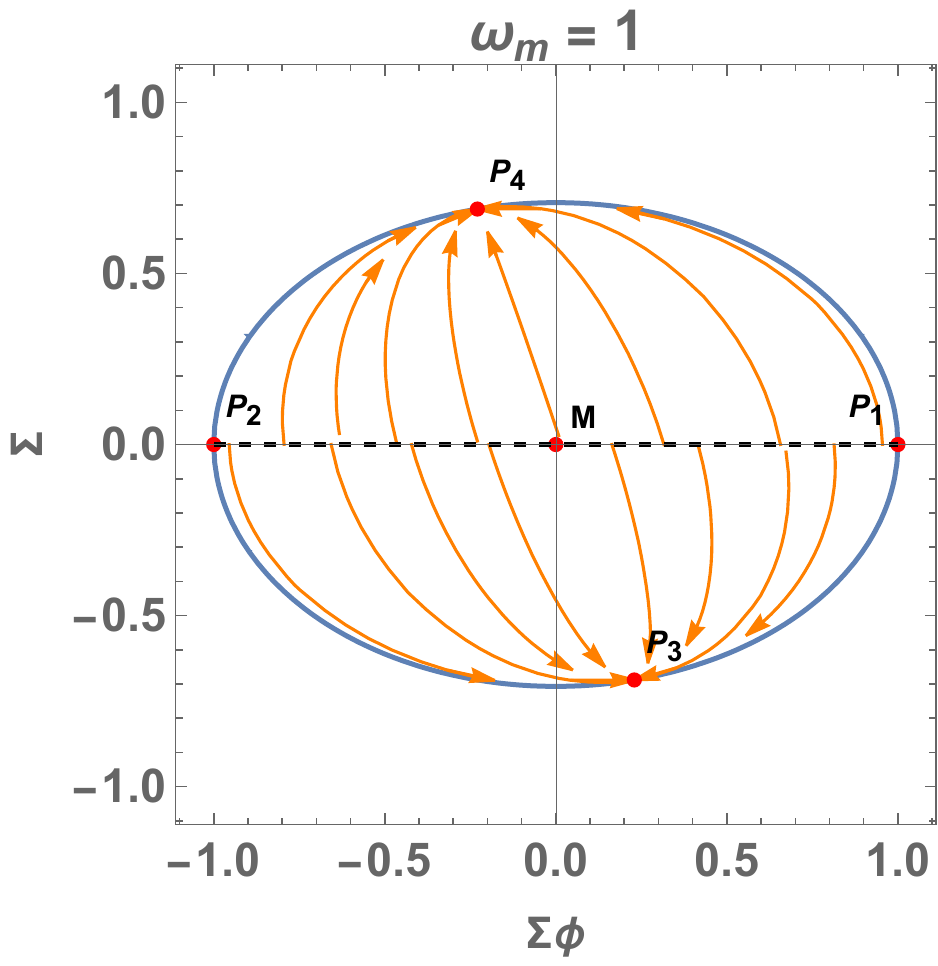}
    \caption{2D projections of the system \eqref{reduced2a}-\eqref{reduced2b} for $\epsilon=+1$, $\mu=2$ and different values of $\omega_m$.}
    \label{fig:4}
\end{figure*}

In Fig. \ref{fig:4} are presented 2D projections of the system \eqref{reduced2a}-\eqref{reduced2b} for $\epsilon=+1$, $\mu=2$ and different values of $\omega_m$. 
   
   \subsubsection{Case $\epsilon=-1$}

The equilibrium points of the system \eqref{reduced2a} and \eqref{reduced2b}  for $\epsilon=-1$ are summarised in Tab. \ref{tab:B1}. 

  \begin{table*}[!t]
    \centering
    \begin{tabular}{|c|c|c|c|}
    \hline
  \text{Label}& \text{Coordinates}\; $(\Sigma_{\phi},  \Sigma)$ & \text{Eigenvalues }& \text{Stability}\\\hline 
  $L_1$&$(\Sigma_{\phi},-3\Sigma_{\phi})$& $\{-3, -3 (\omega_m +1)\}$& \text{Stable}\\ \hline
  $P_6$ & $ \left(\frac{1}{\sqrt{9 \mu -1}}, -\frac{3}{\sqrt{9 \mu -1}} \right)$ & $\{-3, -3 (\omega +1)\}$& \text{Stable} \\\hline
  $P_7$ & $\left(-\frac{1}{\sqrt{9 \mu -1}}, \frac{3}{\sqrt{9 \mu -1}} \right)$ & $\{-3, -3 (\omega +1)\}$& \text{Stable} \\\hline 
  $M$&$(0,0)$& $\left\{\frac{3 (\omega_m -1)}{2}, 3 (\omega_m +1)\right\}$ & \text{Stable for } $\omega=-1$\\&&&\text{Unstable for }$\omega=1$\\&&&\text{Saddle otherwise}\\ \hline
    \end{tabular}
    \caption{Equilibrium points of system \eqref{reduced2a} and \eqref{reduced2b} for $\epsilon=-1$ with their eigenvalues and stability.}
    \label{tab:B1}
\end{table*}
  
\begin{figure*}[ht]
    \centering
     \includegraphics[scale=0.7]{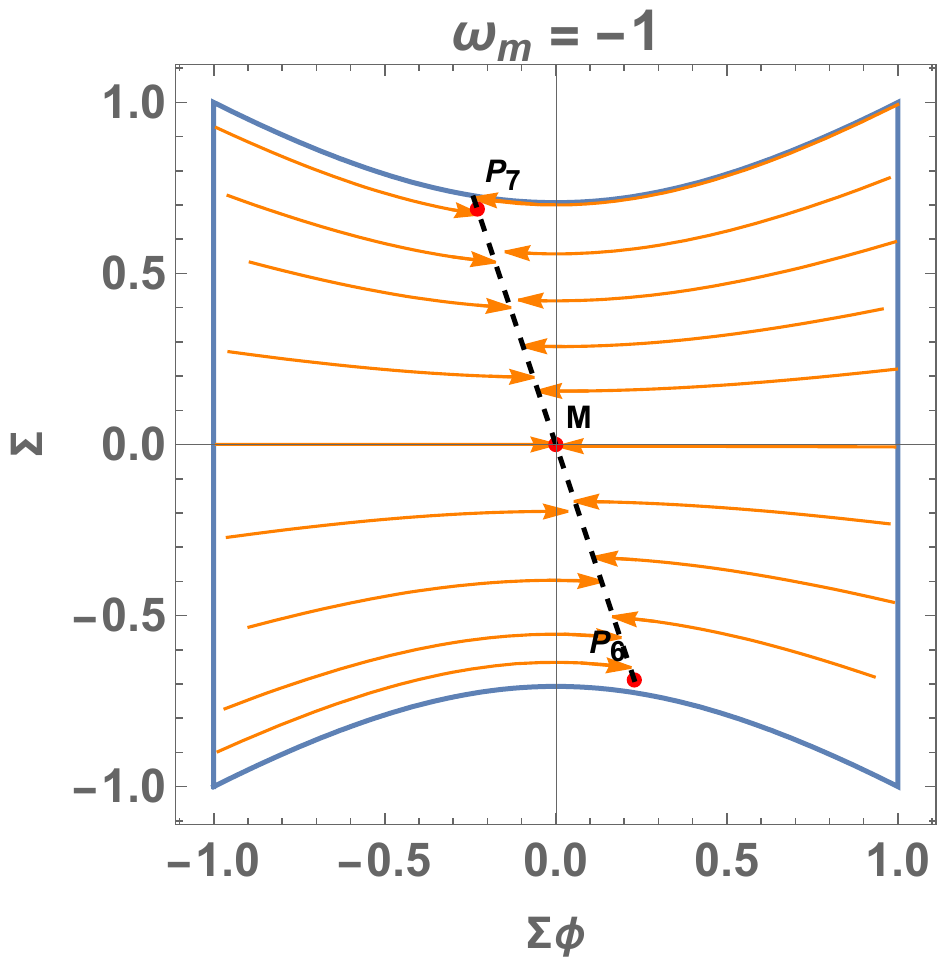}
     \includegraphics[scale=0.7]{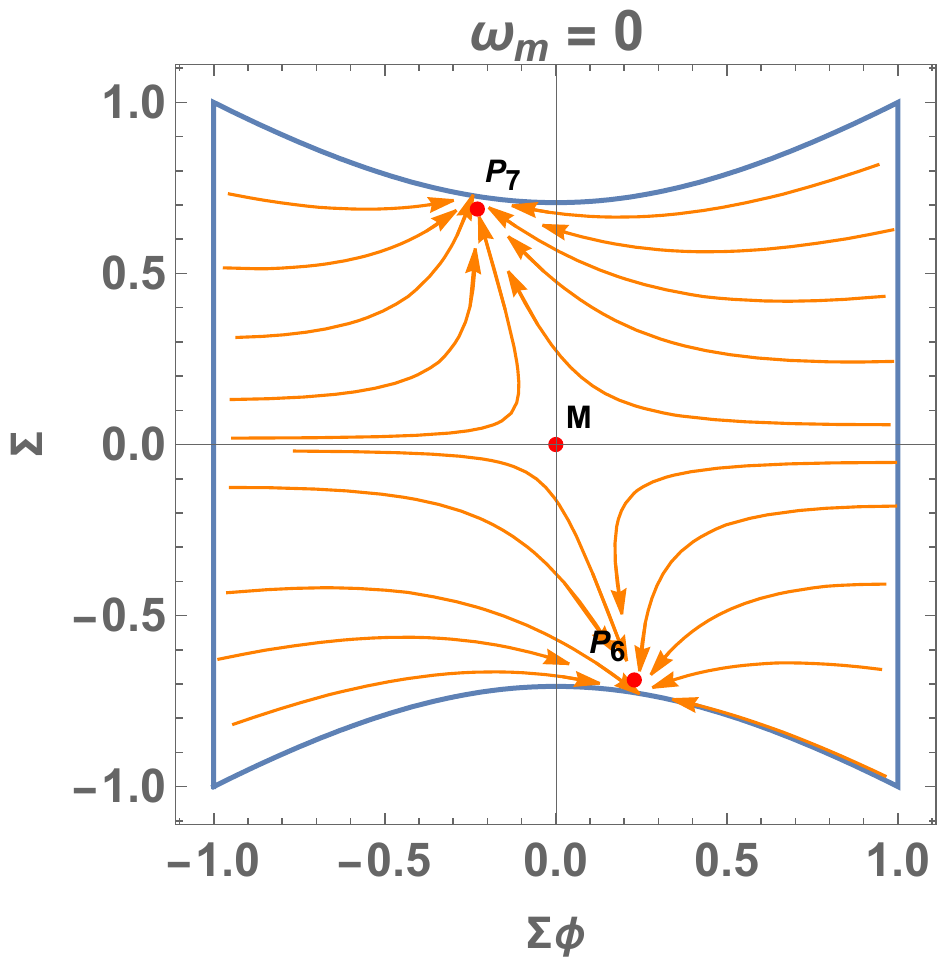}
     \includegraphics[scale=0.7]{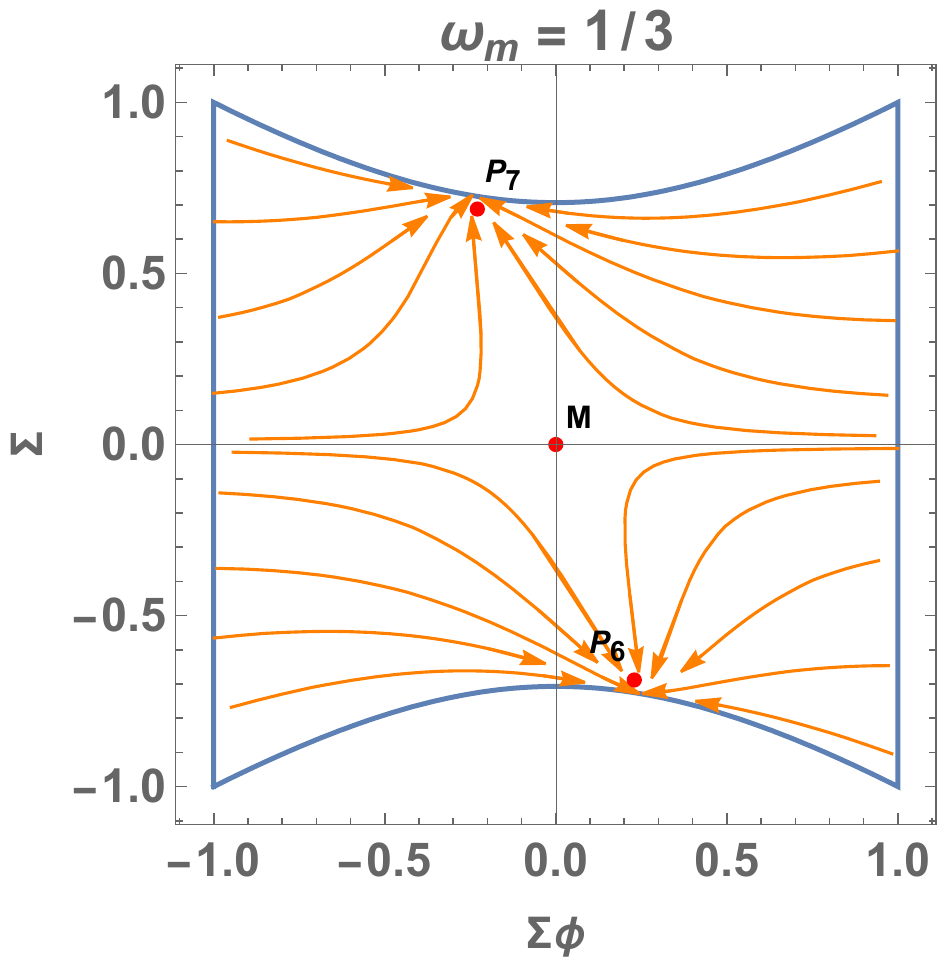}
     \includegraphics[scale=0.7]{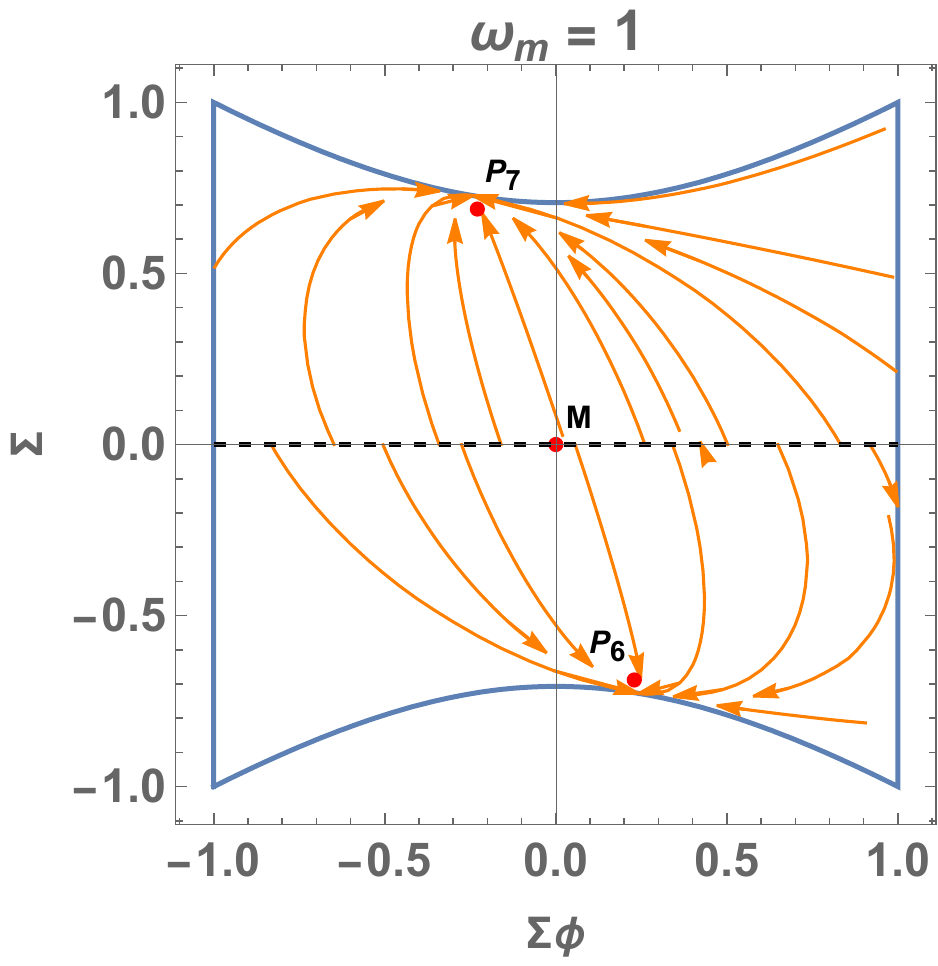}
    \caption{2D projection \eqref{reduced2a}-\eqref{reduced2b} for $\epsilon=-1$, $\mu=2$ and different values of $\omega_m$.}
    \label{fig:6}
\end{figure*}

In Fig. \ref{fig:6} is presented a 2D projection \eqref{reduced2a}-\eqref{reduced2b} for $\epsilon=-1$, $\mu=2$ and different values of $\omega_m$.

\section{Conclusions}
In this study, we investigated the effects of the modification of the Poisson algebra on the dynamics of scalar field cosmology. Specifically, we performed a detailed analysis of the phase-space by studying the equilibrium points and their stability, such that, reconstructing the cosmological history.

The modified Poisson algebra modifies the field equations such that a cosmological constant term is introduced and the pressure component of the scalar field's energy-momentum tensor is different from that of the canonical scalar field. Moreover, a mass term for the scalar field is introduced, which is described by the cosmological constant. 

As a result, the equilibrium points provided by the modified field equations are different from that of the usual scalar field model. From the analysis, we can conclude that the modified equations can provide more than one accelerated universe, described by the de Sitter solution. Hence, cosmic inflation and late-time acceleration are provided by the specific theory.

In the matter-less case, we divided the study into two subcases, one for $\epsilon=1$ and one for $\epsilon=-1$. In total, we have six families of physically accepted equilibrium points which can describe stiff fluid solutions and de Sitter spacetime in the asymptotic regime. 

In the case with the matter, we also consider the subcases $\epsilon=\pm 1,$ and in total, we obtained eight families of equilibrium points, the same ones as in the case without matter and just one additional equilibrium point which describes matter.

In future work, we plan to investigate further the modified field equations with the introduction of a nonzero scalar field potential, while an interacting term between the scalar field and the matter source will be considered.

\section*{Acknowledgments}

Genly Leon was funded by Vicerrectoría de Investigación y Desarrollo Tecnológico (Vridt) at Universidad Católica del Norte (UCN) through
Concurso De Pasantías De Investigación Año 2022, Resolución Vridt N° 040/2022 and through Resolución Vridt N° 054/2022. He also thanks the support of Núcleo de Investigación Geometría Diferencial y Aplicaciones, Resolución Vridt N°096/2022. Andronikos Paliathanasis acknowledges Vridt-UCN through Concurso de Estadías de Investigación, Resolución VRIDT N°098/2022. Alfredo David  Millano was supported by Agencia Nacional de Investigación y Desarrollo - ANID-Subdirección de Capital Humano/Doctorado Nacional/año 2020- folio 21200837.

\end{document}